\documentclass[twocolumn]{aastex631}

\usepackage[T1]{fontenc}
\usepackage{textcomp, gensymb}
\usepackage{amsmath}
\usepackage{upgreek}
\usepackage{subfigure}
\usepackage{orcidlink}

\received{2024 September 28}
\revised{2025 March 25}
\accepted{2025 April 1}
\published{2025 May 8}
\shortauthors{Feng, Yuan, \& Zhang}

\begin{document}

\title{Reflection Spectra of Accretion Disks Illuminated by an Off-Axis Corona}

\author[0000-0003-0554-9425]{Yuan Feng}
\affiliation{School of Astronomy and Space Science, University of Science and Technology of China, Hefei 230026, People's Republic of China}
\affiliation{CAS Key Laboratory for Research in Galaxies and Cosmology, Department of Astronomy, University of Science and Technology of China, Hefei 230026, People's Republic of China}

\author[0000-0002-7330-4756]{Ye-Fei Yuan}
\email{yfyuan@ustc.edu.cn}
\affiliation{School of Astronomy and Space Science, University of Science and Technology of China, Hefei 230026, People's Republic of China}
\affiliation{CAS Key Laboratory for Research in Galaxies and Cosmology, Department of Astronomy, University of Science and Technology of China, Hefei 230026, People's Republic of China}

\author[0000-0001-5586-1017]{Shuang-Nan Zhang}
\email{zhangsn@ihep.ac.cn}
\affiliation{University of Chinese Academy of Sciences, Chinese Academy of Sciences, Beijing 100049, People's Republic of China}
\affiliation{Key Laboratory of Particle Astrophysics, Institute of High Energy Physics, Chinese Academy of Sciences, Beijing 100049, People's Republic of China}

\begin{abstract}
Relativistic reflection features in the X-ray spectra of accreting black holes are considered to be generated by the illumination of the accretion disk by the hot corona. In this work, we present a numerical method for the emission line profile and the reflection spectrum produced by an off-axis X-ray source.
The X-ray source is considered as a point source, as in the lamppost scenario, except that it is located off-axis and moves at arbitrary velocity.
The observed flux for the distant observer is calculated directly without priority evaluation of the emissivity on the accretion disk, which allows our model to be applicable to the point source that deviates from the axis of the black hole spins and moves with a velocity.
To study the impact of the off-axis geometry on the measurement of source properties, we simulate observations for a black hole binary with NuSTAR and eXTP.
We compare the simulation with the observation of the phase-resolved spectra of the low-frequency quasiperiodic oscillation observed by the Insight Hard X-ray Modulation Telescope.
Due to the nonaxisymmetric illumination on the accretion disk, parameters of the model are not reproduced by the lamppost model, including the corona height, radial velocity, and the reflection fraction.
On the other hand, all the model parameters are recovered through the off-axis model.
\end{abstract}

\keywords{\href{http://astrothesaurus.org/uat/1811}{X-ray binary stars (1811)}; \href{http://astrothesaurus.org/uat/159}{Black hole physics (159)}; \href{http://astrothesaurus.org/uat/739}{High energy astrophysics (739)}}

\section{Introduction}
Relativistic reflection features are commonly observed in the X-ray spectra of accreting black holes, including black hole X-ray binaries \citep{1989MNRAS.238..729F, 2008MNRAS.387.1489R, 2013ApJ...775L..45M} and active galactic nuclei \citep{1995Natur.375..659T, 2012MNRAS.422.1914D, 2013Natur.494..449R}.
They are generated by the illumination of the accretion disk surrounding the black hole by the hot corona \citep{1995MNRAS.277L..11F}.
The most prominent feature in the X-ray reflection spectrum is the Fe K$\upalpha$ emission line, which is at 6.4 keV in the rest frame of the emitting material \citep{2005MNRAS.358..211R, 2013ApJ...768..146G}.
Since the reflection emission is from the innermost part of the accretion disk, it is strongly affected by general relativity effects near the black hole, including light bending, gravitational redshift, and Doppler effects from the orbital motion of the accreting material \citep{2000PASP..112.1145F, 2010MNRAS.409.1534D}.
Therefore, the reflection spectra including emission line profiles are broadened and skewed for the distant observer.

There has been a lot of work on fitting X-ray reflection spectra to the theoretical model, in order to study the structure of the inner part of accretion disk \citep{2020PhRvD.101l3014C, 2021ApJ...913..129T}, analyze the coronal properties \citep{2012MNRAS.424.1284W, 2019Natur.565..198K}, and measure the spins of the black hole \citep{2013CQGra..30x4004R, 2023ApJ...946...19D}.
In order to model the relativistic reflection spectrum, it is essential to make assumptions about the illumination pattern of the accretion disk, that is, the emissivity distribution of the disk, which is determined by the location, scale, and motion of the corona \citep{2012MNRAS.424.1284W}.
A widely used scheme in studies of relativistic reflection spectra currently is the lamppost geometry, where a point source on the spin axis above the black hole illuminates the accretion disk \citep{1991A&A...247...25M, 1996MNRAS.282L..53M}.
It naturally explains the observed steep emissivity distribution in the inner part of the accretion disk \citep{2001MNRAS.328L..27W, 2009Natur.459..540F, 2010MNRAS.406.2591P}.
In addition, several models with extended coronal geometries are investigated \citep{2020A&A...641A..89S, 2022ApJ...925...51R}.
What these models have in common is the assumption that the coronal geometry and the emissivity distribution of the accretion disk are axisymmetric.

The observation of quasiperiodic oscillations (QPOs) in black hole X-ray binaries provides strong evidence that these quasiperiodic signals originate from the geometric effect of the accreting material close to the black hole \citep{2016MNRAS.460.2796S, 2021NatAs...5...94M}.
The studies of a large size of statistical samples of QPOs show that the Type-C QPOs could be caused by the precession of the corona \citep{2015MNRAS.448.3348H, 2015MNRAS.447.2059M, 2017MNRAS.464.2643V}.
By applying phase-resolved analysis for the QPOs, the significant modulations of the reflection component with QPO phases are measured from the Type-C QPO of H 1743$-$322 \citep{2016MNRAS.461.1967I, 2017MNRAS.464.2979I}, GRS 1915+105 \citep{2022MNRAS.511..255N}, and MAXI J1820+070 \citep{2023ApJ...957...84S}.
The Insight Hard X-ray Modulation Telescope (HXMT) has a broad energy coverage and a large effective area in the hard X-ray band, which allows for detailed timing analysis on high-energy and broadband variability of bright X-ray sources \citep{2014SPIE.9144E..21Z, 2020SCPMA..6349502Z}.
Using Insight-HXMT observations, the phase-resolved spectroscopy in a broad energy band for low-frequency QPOs (LFQPOs) in black hole binary MAXI J1820+070 during its 2018 outburst is analyzed \citep{2021NatAs...5...94M, 2023ApJ...957...84S}.
The modeling for phase-resolved spectra shows that the illumination profile of the accretion disk tends to be asymmetric rather than axisymmetric, which suggests that the corona near the black hole has a nonaxisymmetric geometry and probably a significant component of the outflow velocity \citep{2012MNRAS.427..934I}.

There is some work to explain the variability of emission lines by calculating the disk emissivity and emission line profile produced by precessing inner flows \citep{2012MNRAS.427..934I, 2020ApJ...897...27Y} or by an off-axis point source \citep{2000MNRAS.311..161Y, 2012MNRAS.424.1284W, 2023ApJ...955...53F}.
Most of these studies use the Monte Carlo method, that is too expensive and makes models difficult to fit the observational data directly.

In this paper, we present a numerical method for calculating the reflection spectrum produced by the illumination by an off-axis point source.
By introducing the lensing factor, the observed emission line profiles are calculated directly by integration, and the priority evaluation of the emissivity on the accretion disk is avoided, which allows our method to be applicable to the point source which deviates from the axis of the black hole spins and moves with a velocity.
In addition, our method has fast computational efficiency and simple estimation of errors.
In Section \ref{Theory}, we present the theoretical derivation of our method. In Section \ref{Numerical Model}, we estimate the errors of the numerical models for emission line profiles and reflection spectra. The effects of the off-axis corona on the observational data are presented in Section \ref{Effect in Observation}.
We discuss our results in Section \ref{Discussion} and summarize them in Section \ref{Conclusion}.

\section{Theory}
\label{Theory}
\subsection{Off-Axis Corona}
\begin{figure*}[htb!]
    \centering
    \includegraphics[width=0.85\textwidth]{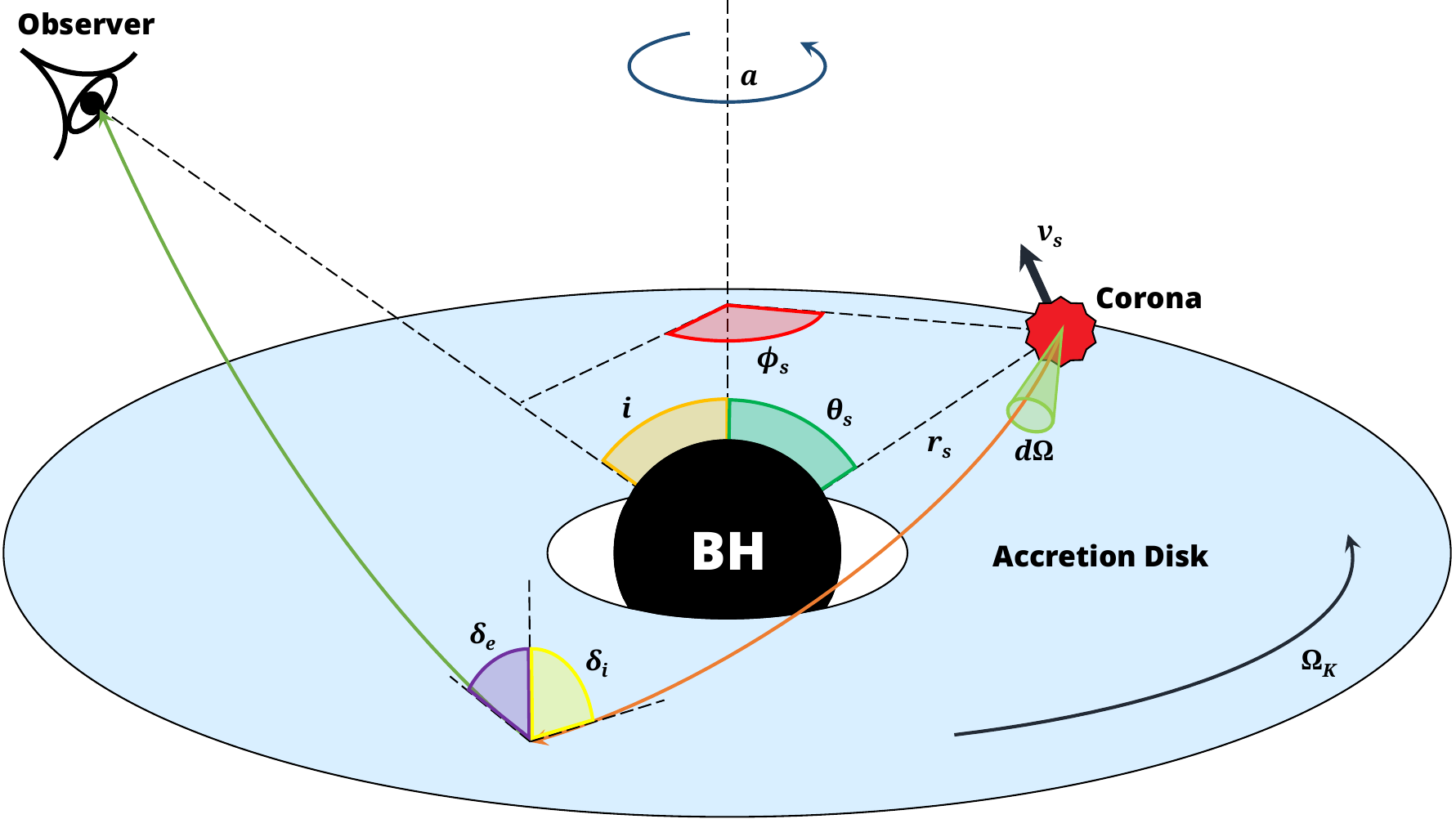}
    \caption{The schematic of off-axis corona model. The corona is simplified as an off-axis point source with a velocity, illuminating the accretion disk near the black hole. The photons emitted from the corona are reprocessed and reflected in the disk. The emitted photons observed by the distant observer, producing the observed reflection spectrum.}
    \label{model}
\end{figure*}

The line element of Kerr metric in the Boyer-Lindquist coordinates $(t,~r,~\theta,~\phi)$ can be written as \citep{1972ApJ...178..347B}
\begin{equation}
    ds^2 = -e^{2\nu} dt^2 + e^{2\psi} (d\phi - \omega dt)^2 + e^{2\mu_1} dr^2 + e^{2\mu_2} d\theta^2,
\end{equation}
where
\begin{equation}
    \begin{aligned}
         & e^{2\nu} = \Sigma\Delta / \mathcal{A},\qquad e^{2\psi} = \sin^2\theta \mathcal{A} / \Sigma, \\
         & e^{2\mu_1} = \Sigma / \Delta,\qquad e^{2\mu_2} = \Sigma,\qquad \omega = 2ar / \mathcal{A},
    \end{aligned}
\end{equation}
and
\begin{equation}
    \begin{aligned}
         & \Delta = r^2 - 2r + a^2,                               \\
         & \Sigma = r^2 + a^2 \cos^2\theta,                       \\
         & \mathcal{A} = (r^2 + a^2)^2 - a^2 \Delta \sin^2\theta.
    \end{aligned}
\end{equation}
The natural units $G = c = 1$ are used, and the mass of the black hole is also taken as $M = 1$ in this paper.

The schematic of our model is shown in Figure \ref{model}.
The accretion disk is assumed to be geometrically thin and optically thick.
The disk material orbits in Keplerian circular motion around the central black hole in the equatorial plane.
The distant observer is assumed to be located at $(r_\infty,~i,~0\degree)$, where the coordinate radius is set to $r_\infty = 10^{11}$ in this work.
The X-ray corona is assumed to be a point source located at $(r_s,~\theta_s,~\phi_s)$.
The physical velocities of the corona $(v_r,~v_\theta,~v_\phi)$ with respect to the locally nonrotating reference frame can be written as \citep{1972ApJ...178..347B}
\begin{equation}
    v_r = e^{\mu_1 - \nu} \dot{r},\quad v_\theta = e^{\mu_2 - \nu} \dot{\theta},\quad v_\phi = e^{\psi - \nu} (\dot{\phi} - \omega),
\end{equation}
where $\dot{r} = dr_s / dt_s,~\dot{\theta} = d\theta_s / dt_s,$ and $\dot{\phi} = d\phi_s / dt_s$ are the coordinate velocities of the corona.

\subsection{Illumination}
The photons are emitted isotropically in the rest frame of the corona.
A photon from the solid angle $d\Omega$ in the rest frame of the corona has a four momentum $p'_{(\alpha)}$.
By tracing the trajectory of this photon, the position on the accretion disk hit by the photon $(r,~\phi)$ and the energy shift factor from the corona to the disk $g_s = E_\mathrm{disk} / E_s$ can be calculated.
Assuming a power-law shape of the emitted radiation $N \propto E^{-\Gamma}$, the incident flux on the accretion disk can be calculated as \citep{2000MNRAS.315....1R, 2013MNRAS.430.1694D}
\begin{equation}
    F_i(r,~\phi) = g_s^\Gamma \frac{d\Omega}{\gamma^{(\phi)} dA} = g_s^\Gamma \frac{e^\nu}{\gamma^{(\phi)}} \frac{d\Omega}{r dr d\phi},
\end{equation}
where
\begin{equation}
    dA = e^{-\nu} rdrd\phi
\end{equation}
is the area element at $(r,~\phi)$ on the disk.
$\gamma^{(\phi)}$ is the Lorentz factor of the disk material orbiting in Keplerian motion \citep{1972ApJ...178..347B}.

\subsection{Lensing Factor and Line Profile}
The observed intensity can be written as \citep{1966AnPhy..37..487L}
\begin{equation}
    I_\mathrm{obs} (E) = \int \Big(\frac{E}{E_e}\Big)^3 I_e(E_e) d\alpha d\beta,
\end{equation}
where the emission intensity originating from the disk is assumed as monoenergetic $I_e (E_e) = \varepsilon (r,~\phi) \delta (E_e - E_0)$ \citep{2010MNRAS.409.1534D} and the disk emissivity is proportional to the incident flux from the corona $\varepsilon \propto F_i$.
The impact parameters $\alpha$ and $\beta$ are the coordinates of the impact position of a photon on the plane of the observer \citep{1973ApJ...183..237C, 1975ApJ...202..788C}.

The lensing factor $\ell$ is defined as the ratio of the area subtended by photons at infinity perpendicular to light rays through which photons arrive at the observer, to the disk area from where these photons originate \citep{2004ApJS..153..205D}, which can be written as
\begin{equation}
    \ell = \frac{g_\mathrm{obs}}{\cos \delta_e} \frac{d\alpha d\beta}{rdrd\phi},
\end{equation}
where $g_\mathrm{obs} = E / E_e$ is the energy shift factor from the accretion disk to the observer and $\delta_e$ is the emission angle, which is the angle between the emitted light and the normal of the disk.

By using the lensing factor, the observed intensity can be rewritten as
\begin{equation}
    I_\mathrm{obs} (E) = \int g_\mathrm{obs}^2 g_s^\Gamma \frac{e^\nu}{\gamma^{(\phi)}} \delta (E_e - E_0) \ell \cos \delta_e d\Omega.
\end{equation}
In observation, the contribution of the intensity in each energy bin $E_\mathrm{lo} \le E_i < E_\mathrm{hi}$ is more important.
Integrating the equation over the energy gives the observed flux within one energy bin
\begin{equation}
    \int_{E_\mathrm{lo}}^{E_\mathrm{hi}} I_\mathrm{obs} dE = \int\limits_{E_\mathrm{lo} \le g_\mathrm{obs} E_0 < E_\mathrm{hi}} g_\mathrm{obs}^3 g_s^\Gamma \frac{e^\nu}{\gamma^{(\phi)}} \ell \cos\delta_e d\Omega.
\end{equation}
The integration is over the solid angle in the rest frame of the corona.

\section{Numerical Model}
\label{Numerical Model}
\subsection{\textsl{\texttt{offaxis}}\footnote{The source code of \texttt{offaxis} model can be downloaded from \href{https://github.com/feng1m8/offaxis}{https://github.com/feng1m8/offaxis}.} Model}
The calculation of the observed intensity is divided into two steps.
For a photon with a given four momentum in the rest frame of the corona $p'_{(\alpha)}$, the location where this photon hits the accretion disk is calculated first using the ray-tracing method.
In this work, the ray-tracing code \texttt{YNOGK} \citep{2013ApJS..207....6Y} is used to calculate the photon trajectories and the energy shift from the corona to the accretion disk.
Ray trajectories are calculated using elliptic integrals, which is efficient and accurate in the calculation of the emission from the accretion disk system \citep{1998NewA....3..647C, 2009ApJ...696.1616D, 2009ApJ...699..722Y}.
The second step is the calculation of the information received by the distant observer about the photons from the accretion disk, including the energy shift factor $g_\mathrm{obs}$, the emission angle $\delta_e$, and the lensing factor $\ell$, which are precomputed and stored as a data table available in the relativistic spectral model \texttt{KYN} \citep{2004ApJS..153..205D}.

There are several methods to generate isotropically emitted photons in the rest frame of the X-ray source.
The Monte Carlo method can be used to generate the emitted direction of photons randomly, but this method usually requires a number of calculations.
In this work, the numerical method \texttt{HEALPix} is used to generate the initial momentum of photons \citep{2005ApJ...622..759G}.
The \texttt{HEALPix} method is originally intended for satellite missions to measure the anisotropy of the cosmic microwave background \citep{2003ApJS..148....1B, 2011A&A...536A...1P}, and later it is applied to more fields in data processing and visualization.
This method allows calculations to be performed efficiently at high resolution, and each generated photon in the rest frame of the corona has the same size of solid angle.
The efficiency will be improved if an adaptive method is used; however, no numerical library is found out of the box.
For each photon generated, its trajectory is traced, and
if the trajectory intersects with the surface of the accretion disk, the contribution to the observed flux of this photon is calculated and added to an energy bin of the distant observer.

\begin{figure*}[htb!]
    \centering
    \includegraphics[width=0.85\textwidth]{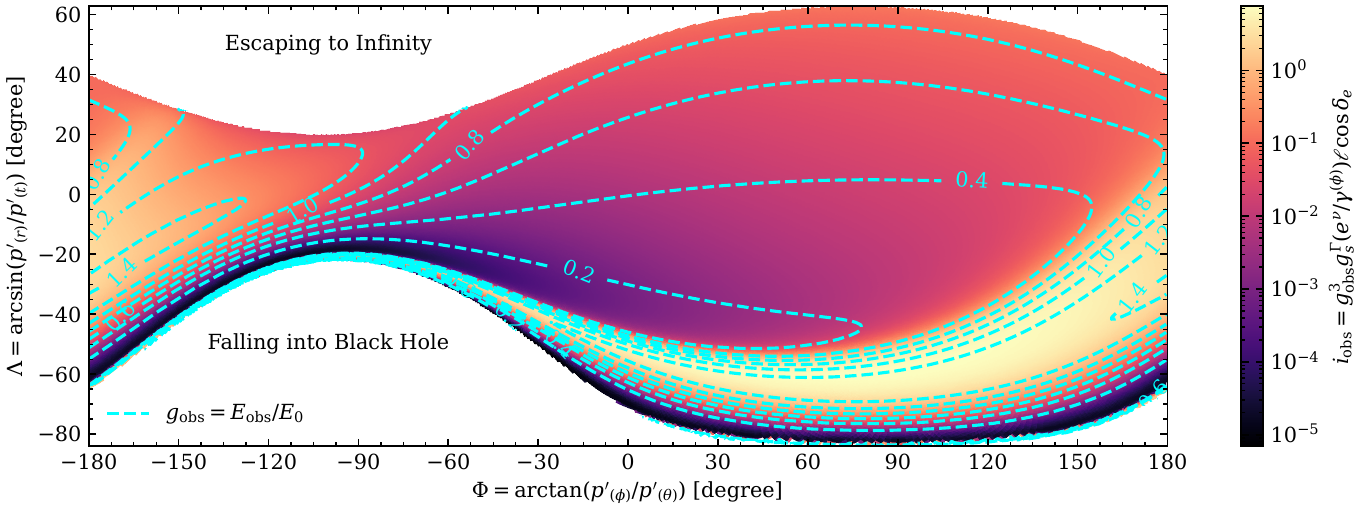}
    \caption{The contribution of photons emitted in different directions in the rest frame of the corona to the observed intensity $i_\mathrm{obs} = g_\mathrm{obs}^3 g_s^\Gamma (e^\nu / \gamma^{(\phi)}) \ell \cos \delta_e$ (filled colors). The dashed lines represent the energy shift factor $g_\mathrm{obs}$ from the disk region where the photons hit the distant observer. $p'_{(\alpha)} = (p'_{(t)},~p'_{(r)},~p'_{(\theta)},~p'_{(\phi)})$ is the four momentum of the emitted photon in the rest frame of the corona. The position of the corona $(r_s,~\theta_s,~\phi_s)$ is set to $(3,~10\degree,~0\degree)$. The spin of the black hole is $a = 0.998$. The observing inclination is $i = 80\degree$, and the photon index of the continuum is $\Gamma = 2$.}
    \label{orbiting_10}
\end{figure*}

\begin{figure*}[htb!]
    \centering
    \includegraphics[width=0.85\textwidth]{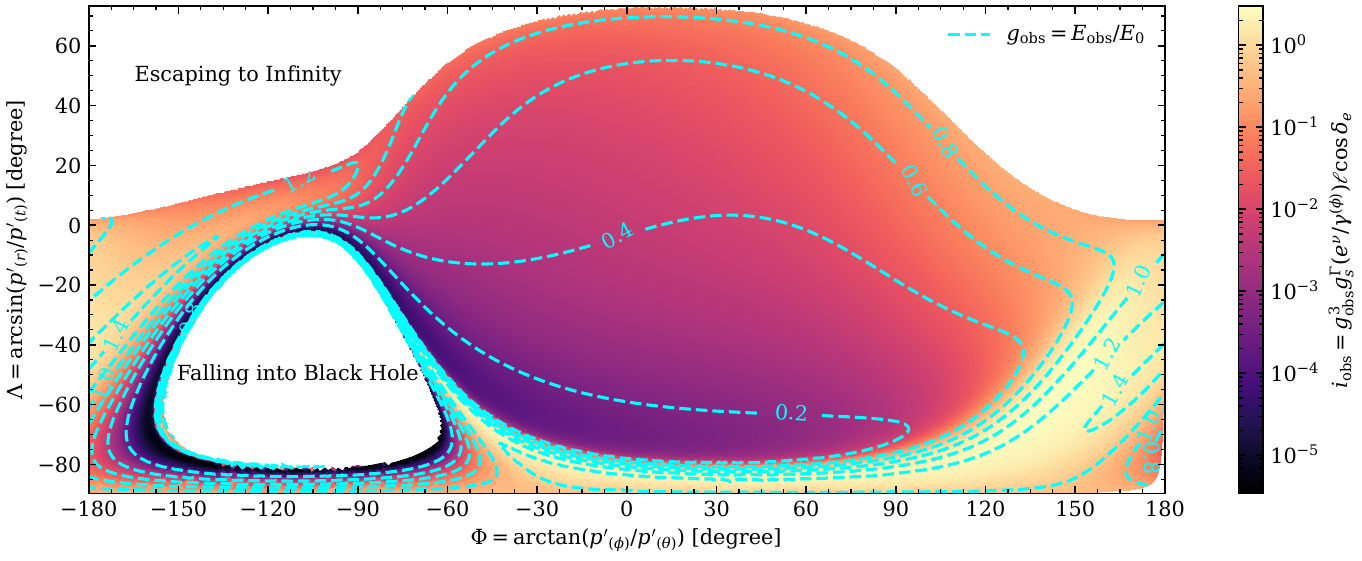}
    \caption{Same as Figure \ref{orbiting_10}, except the inclination of the corona $\theta_s = 60\degree$.}
    \label{orbiting}
\end{figure*}

\begin{figure}[htb!]
    \centering
    \includegraphics[width=0.425\textwidth]{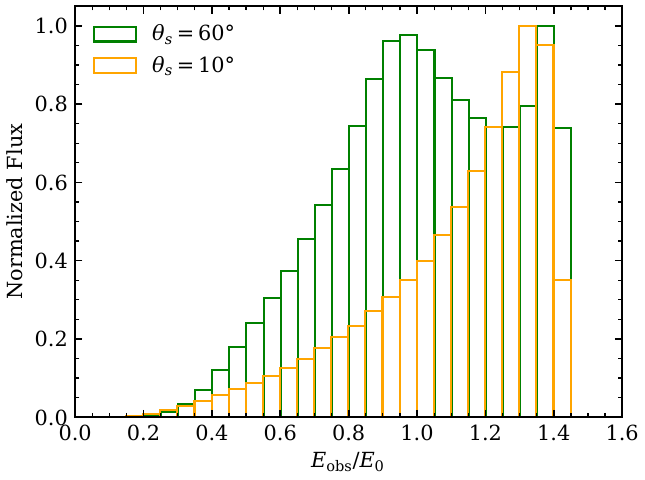}
    \caption{The flux received by the energy bins of the distant observer in the case of Figure \ref{orbiting_10} and \ref{orbiting}. The range of each energy bin is 0.05 $E_0$.}
    \label{flux}
\end{figure}

As an example, we consider the emission line profile produced by an orbiting corona, which is orbiting the axis of the black hole spin above the plane of the accretion disk, corotating with the element of the disk at the same radius, as defined in \cite{2012MNRAS.424.1284W}.
The contribution of photons emitted in different directions in the rest frame of the corona to the observed intensity $i_\mathrm{obs}$ and the energy shift factor from the disk region where the photons hit the distant observer $g_\mathrm{obs}$ are shown in Figure \ref{orbiting_10} and \ref{orbiting}.
The trajectories of photons emitted from the corona are divided into three types based on the final arrival of the photons: escaping to infinity, hitting the accretion disk, or falling into the black hole.
In the case of the axial source, the types of the trajectories are only dependent of the radial component of the four momentum of photons.
That is, in the rest frame of the corona the radius of photon arrival is determined by latitude angle $\Lambda$.
When the corona is off-axis and moving, the $\theta$ and $\phi$ components of the four momentum of the photon have effects.
When the inclination of the corona is small, the distribution of where photons finally arrive is similar to that of an axial source.
Photons with large radial component of momentum escape to infinity, and photons with large negative radial component of momentum fall into the black hole.
When the inclination of the corona is large, photons emitted in the direction opposite to the black hole spin are more likely to fall into the black hole due to the frame dragging effect, while photons with the four momentum $p'_{(r)} / p'_{(t)} = -1$ hit the accretion disk.

For each photon hitting the accretion disk, its contribution to the observed flux $i_\mathrm{obs}\Delta\Omega$ is counted into a certain energy bin $E_\mathrm{lo} \le g_\mathrm{obs} E_0 < E_\mathrm{hi}$ according to the energy shift factor $g_\mathrm{obs}$ from the disk location hit by this photon to the distant observer.
The flux received by the energy bins of the distant observer is shown in Figure \ref{flux}.
There are two peaks in the line profile when the inclination of the corona is large, with energies $E_\mathrm{obs}$ about 1.0 and 1.4 $E_0$ respectively.
The contribution of photons emitted from the corona in different directions to the observed flux is extremely unbalanced.
Photons emitted with longitude angle $\Phi$ between $-60\degree$ and 120\degree, shown in the center region of Figure \ref{orbiting}, contribute little of the observed flux with energy $E_\mathrm{obs}$ below 0.8 $E_0$.
Photons emitted from a small fraction of the radiation directions contribute the majority of the observed flux due to the strong gravitational redshift, which makes the energy shift term $g_s^\Gamma$ dominant \citep{2013MNRAS.430.1694D}.

The relativistic reflection spectra are calculated by the convolution of the monoenergetic line model and the local reflection model of the accretion disk \texttt{xillver} \citep{2010ApJ...718..695G, 2011ApJ...731..131G, 2013ApJ...768..146G}.
The contribution of the emission angle is correctly treated according to \cite{2014ApJ...782...76G}.
The effect of the energy shift of the cutoff energy of the spectrum from the corona to the different region of the accretion disk is also considered.
There are four models provided in total in this work, namely, a line model and its convolution model and two reflection models, of which the profiles of the incident spectra are high-energy cutoff power-law and Comptonization continuum, respectively. The new models are available for XSPEC \citep{1996ASPC..101...17A} and can be publicly downloaded.

\subsection{Error Estimation}
\begin{figure}[htb!]
    \centering
    \subfigure[]{\includegraphics[width=0.425\textwidth]{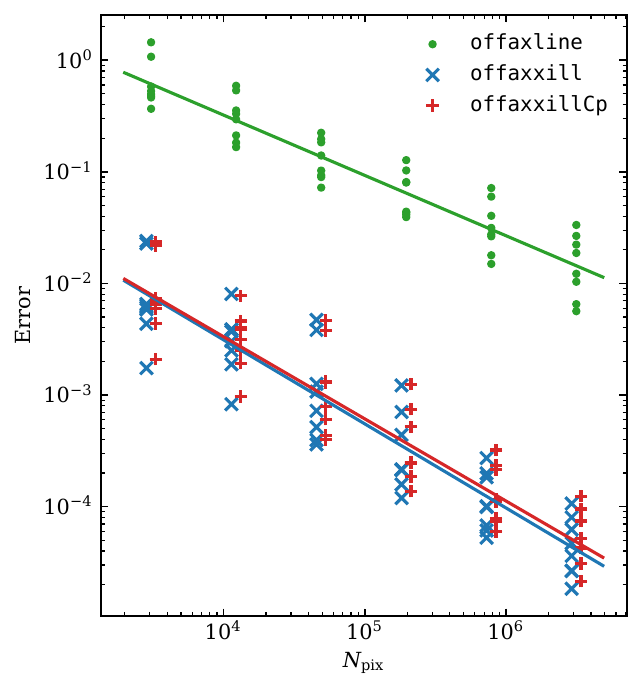}}
    \subfigure[]{\includegraphics[width=0.425\textwidth]{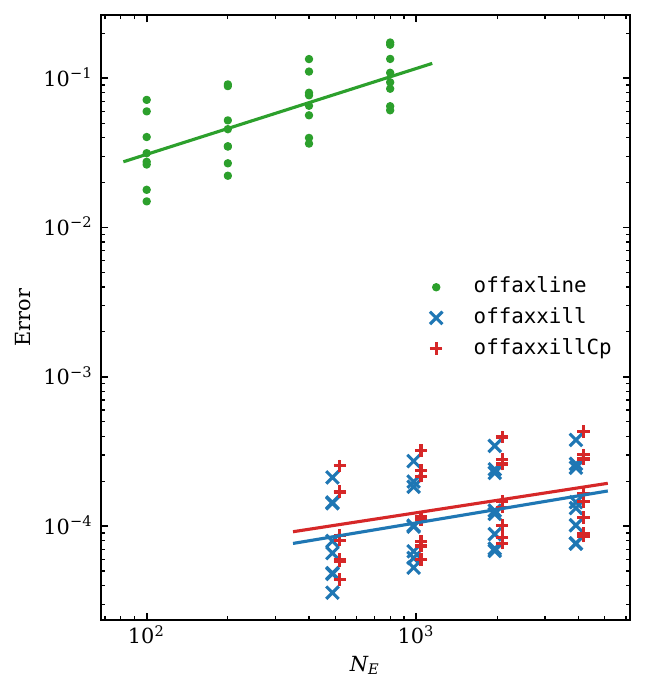}}
    \caption{The relative error of models with different (a) number of divisions $N_\mathrm{pix}$ and (b) number of energy bins $N_E$ in the case of Figure \ref{orbiting}, expect the azimuth angles of the corona $\phi_s$ are set from 0\degree{} to 315\degree, increasing by 45\degree. The results in all models with the number of divisions $12\times 1024^2$ are regarded as the exact results. The number of energy bins $N_E$ is fixed to 100 for the line model and 1000 for the reflection model in subfigure (a). The number of divisions is fixed to $12\times 256^2$ in subfigure (b).}
    \label{err}
\end{figure}

The numerical error of the model is related to the number of energy bins and the number of divisions of the integrating region.
Figure \ref{err} shows the relative errors with different number of divisions and energy bins, where the relative error of the model is measured by the quadratic mean of the relative error in each energy bin.
Among the three models, the error of the line model is obviously larger than that of the reflection model, while the error of \texttt{offaxxillCp} model is slightly larger than that of \texttt{offaxxill} model.
The errors of these models have a power-law relationship with the number of energy bins and divisions approximately, which can be expressed as
\begin{equation}
    e_\mathrm{line} \sim 10^{-2} \Big(\frac{N_E}{10^2}\Big)^{0.6} \Big(\frac{N_\mathrm{pix}}{10^6}\Big)^{-0.5}
\end{equation}
for the line model and
\begin{equation}
    e_\mathrm{refl} \sim 10^{-4} \Big(\frac{N_E}{10^3}\Big)^{0.3} \Big(\frac{N_\mathrm{pix}}{10^6}\Big)^{-0.7}
\end{equation}
for reflection models.

The number of energy bins is usually certain for an observed spectrum.
The appropriate number of divisions is chosen to achieve high computational efficiency while satisfying numerical accuracy.

\section{Effect in Observation}
\label{Effect in Observation}
\subsection{Line Profiles and Reflection Spectra}
\begin{figure*}[htb!]
    \centering
    \includegraphics[width=0.85\textwidth]{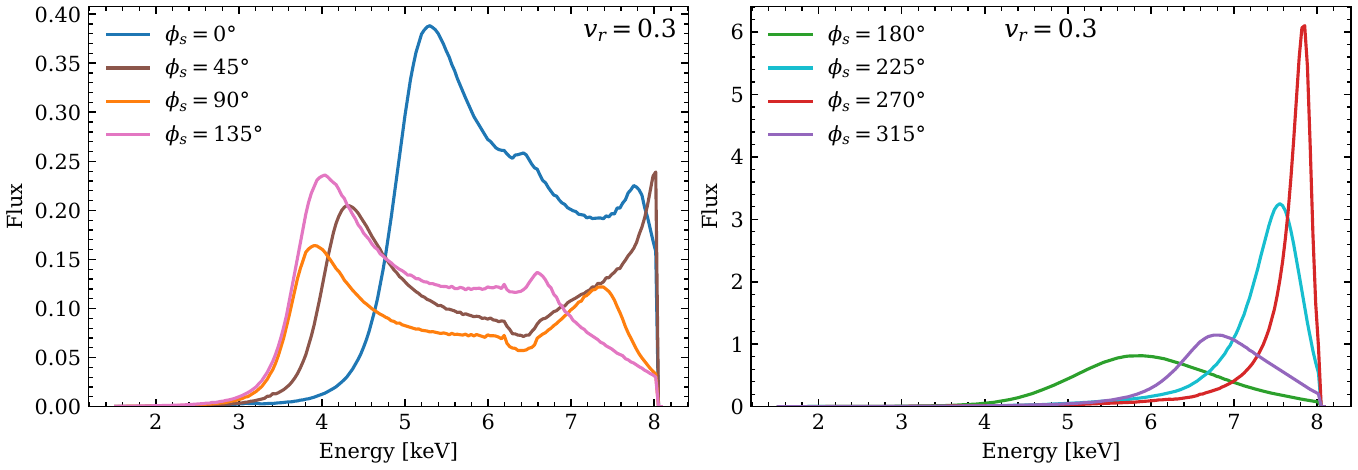}
    \caption{Emission line profiles for different azimuths of the corona. The velocity of the corona is radially outward $v_r = 0.3$. The radius of the corona is $r_s = 7$, and the inclination of the corona is $\theta_s = 75\degree$. The spin of the black hole is $a = 0.998$. The observing inclination is $i = 65\degree$. The photon index of the continuum is $\Gamma = 2$.}
    \label{line}
\end{figure*}

\begin{figure*}[htb!]
    \centering
    \includegraphics[width=0.85\textwidth]{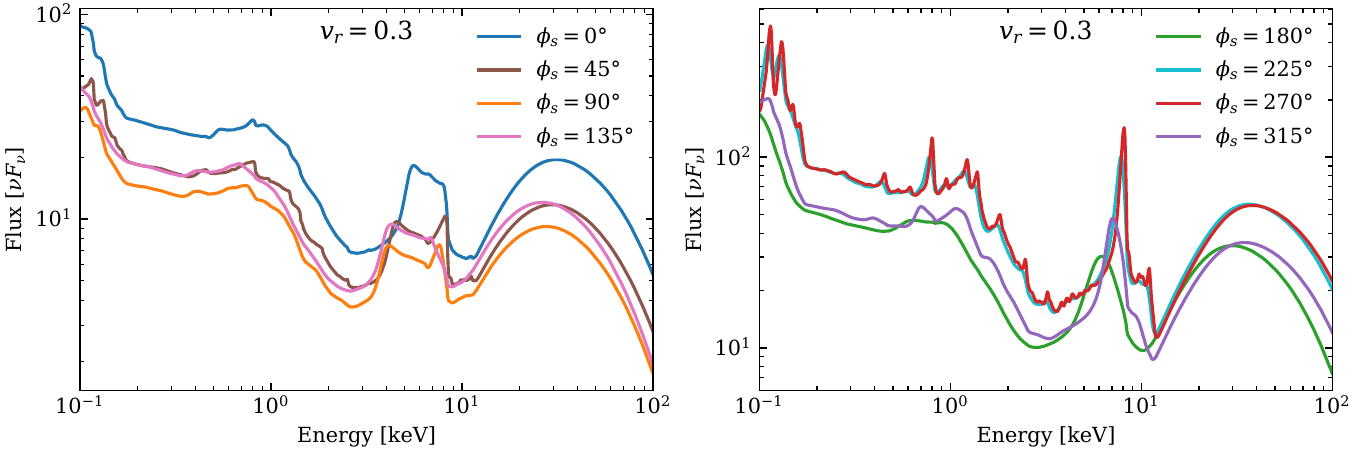}
    \caption{Reflection spectra for different azimuths of the corona corresponding to the case in Figure \ref{line}. The ionization and density of the accretion disk is $\log\xi = 3.0$ and $N = 10^{15}$ cm$^{-3}$. The electron temperature of the corona is $kT_e = 60$ keV. The iron abundance is $A_\mathrm{Fe} = 5.0$ (with respect to solar abundances).}
    \label{reflection}
\end{figure*}

The illumination and emission profile of the accretion disk is dependent on the position and velocity of the corona, as the disk region below the corona receives more flux, which is the most important factor affecting the line profiles and reflection spectra \citep{2000MNRAS.311..161Y, 2012MNRAS.424.1284W, 2023ApJ...955...53F}.
In the lamppost model, observed emission lines are broad and skewed with an extended red shifted wing and a sharp blue shifted peak on profiles \citep{2000PASP..112.1145F, 2021ARA&A..59..117R}.
When the height of the corona is large, the line profile becomes narrow, with a double-peaked structure in which the red peak is weaker than the blue peak \citep{2013MNRAS.430.1694D}.
Our model regresses well to the lamppost model at the inclination of the corona $\theta_s = 0\degree$.
While allowing the corona deviating from the axis of the black hole spin, the emission lines exhibit different properties.
As two examples, we show line profiles and reflection spectra produced by the off-axis corona with different velocity directions, which are moving radially outward or orbiting around the axis of the black hole spin with a certain velocity $v_r = 0.3$ or $v_\phi = 0.3$, respectively.
Figure \ref{line} shows the emission line profiles for different azimuths of the corona, of which the velocity is radially outward.
When the corona is above the receding side of the accretion disk ($0\degree < \phi_s < 180\degree$), photons emitted from the disk are redshifted more. The red peak of the line profile is stronger than the blue peak.
In the case of $\phi_s = 135\degree$ in Figure \ref{line}, the blue peak is too weak to distinguish in the line profile.
When the corona is above the approaching side of the accretion disk ($180\degree < \phi_s < 360\degree$), photons emitted from the disk are more observed blue shifted.
There is a strong peak on the line profile, while the red part is weak, and it is difficult to observe the red wing in the profile.
A special case is that when the coronal azimuth $\phi_s = 180\degree$, the peak energy of the emission line is lower than the energy of the radiation photon in its local reference frame, i.e. the energy of the iron emission line is observed to be lower than 6.4 keV, which cannot be explained by the lamppost model.

Due to the nonaxisymmetric illumination from the off-axis corona, there are large differences of the total flux and the shape of the reflection spectra at different azimuths of the corona, as shown in Figure \ref{reflection}.
The intensity changes of the iron emission line and Compton hump reach more than 1 order of magnitude.
When the corona is above the receding side of the disk, the features in reflection spectra are relativistically blurred.
There are no obvious narrow lines in the reflection spectrum, and the emission lines are broadened to a large energy range.
In the case of $\phi_s = 45\degree$ and 90\degree, the iron emission line is weakly peaked.
When the corona is above the approaching side of the disk, the emission lines on the reflection spectrum are narrowed.
When the azimuth of the corona is between 90\degree{} and 180\degree, the peak energy of the Compton hump is low due to the strong relativistic redshift effect.
The peak energy of the Compton hump is less than 20 keV when the azimuth of the corona $\phi_s = 135\degree$.

\begin{figure*}[htb!]
    \centering
    \includegraphics[width=0.85\textwidth]{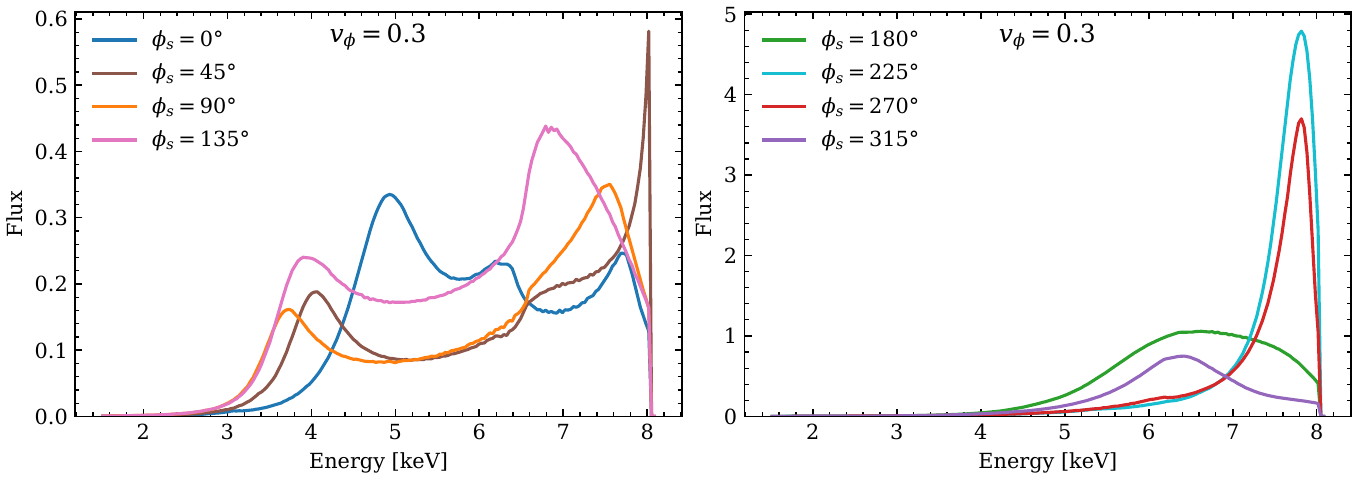}
    \caption{Emission line profiles for different azimuths of the corona. The parameters in the calculation are the same as those in Figure \ref{line}, except that the velocity of the corona is along the azimuthal direction $v_\phi = 0.3$.}
    \label{line_phi}
\end{figure*}

\begin{figure*}[htb!]
    \centering
    \includegraphics[width=0.85\textwidth]{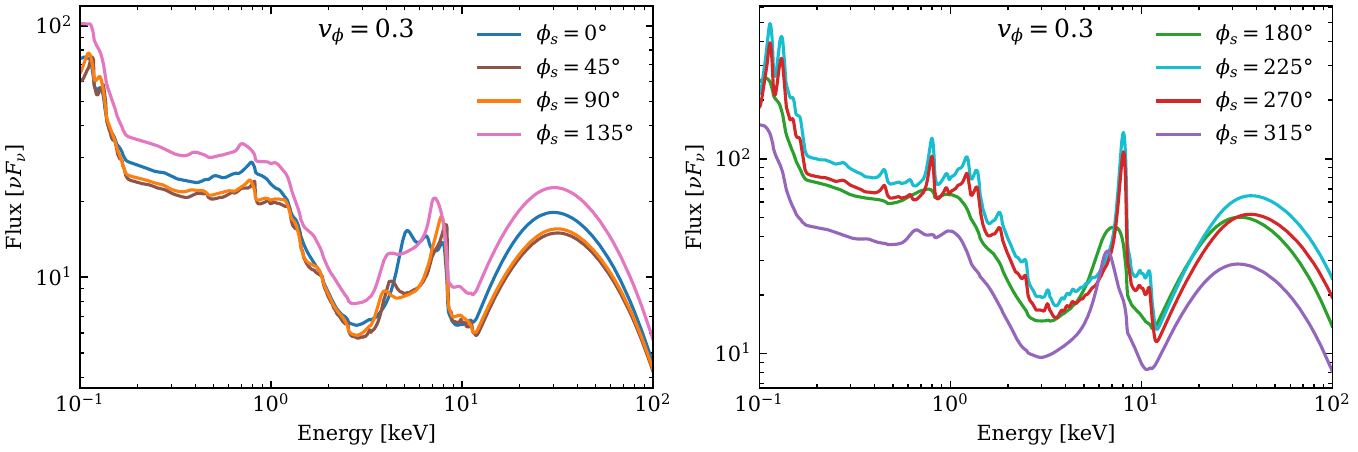}
    \caption{Reflection spectra for different azimuths of the corona corresponding to the case in Figure \ref{line_phi}. The ionization and density of the accretion disk are $\log\xi = 3.0$ and $N = 10^{15}$ cm$^{-3}$. The electron temperature of the corona is $kT_e = 60$ keV. The iron abundance is $A_\mathrm{Fe} = 5.0$ (with respect to solar abundances).}
    \label{reflection_phi}
\end{figure*}

Figure \ref{line_phi} and \ref{reflection_phi} show the line profiles and reflection spectra for different azimuths of the corona, of which the velocity is along the azimuthal direction.
Different with the emission lines produced by the corona moving radially, those with the corona moving in the azimuthal direction have the blue peak in the profiles.
In the case of $\phi_s = 45\degree$, 90\degree, and 135\degree, there are two peaks on line profiles.
In the case of $\phi_s = 0\degree$, there are three peaks on the profile, with the additional weak peak at about 6.4 keV, as shown in Figure \ref{reflection_phi}.

\subsection{Simulated Observations}
In order to investigate the effects of off-axis corona geometries on the measurements of source properties in observations,
we consider the case of a bright black hole X-ray binary with the average flux in the energy range from 1 to 10 keV of $10^{-8}$ erg s$^{-1}$ cm$^{-2}$.
We simulate the observational spectra with NuSTAR \citep{2013ApJ...770..103H} and eXTP \citep{2016SPIE.9905E..1QZ}.
The black hole spin and observing inclination are taken to be $a = 0.14$ and $i = 63\degree$ respectively, which are the same as those obtained on MAXI J1820+070 \citep{2020MNRAS.493L..81A, 2021MNRAS.504.2168G, 2021ApJ...916..108Z}.

The corona is assumed to deviate from the axis of the black hole spin and move radially outward.
The radius and inclination of the corona are set to $r_s = 7$ and $\theta_s = 15\degree$, respectively.
The radial velocity of the corona is set to $v_r = 0.5$ with respect to the locally nonrotating reference frame.
In observation, the measured velocities of the corona are typically relative to the corotating reference frame of the corona, which has a relative velocity in the azimuth direction $v_\phi = e^{\psi-\nu} (\Omega - \omega)$ to the locally nonrotating reference frame.
In the case of precessing jets around a stellar mass black hole, the relative velocity is usually about $v_\phi \sim 10^{-4}$, significantly less than the radial velocity of the corona $v_r$.
Therefore, the velocity of the corona in different frames has little effect on measurements.

The photon index of the continuum is set to $\Gamma = 1.6$, and the electron temperature of the corona is $kT_e = 60$ keV.
The inner radius of the accretion disk is set to the radius of the innermost stable circular orbit $R_\mathrm{in} = r_\mathrm{ms} = 5.535$, and the outer radius is fixed to $R_\mathrm{out} = 1000$.
The ionization, iron abundance, and density of accretion disk are set to $\log\xi = 1.0$, $A_\mathrm{Fe} = 5.0$ (with respect to solar abundances) and $N = 10^{15}$ cm$^{-3}$, respectively, in order to eliminate the effects of the properties of the accretion disk on fitting.
The \texttt{Tbabs} component is added to account for interstellar absorption.
The equivalent hydrogen column $N_\mathrm{H}$ is fixed at $0.15\times 10^{22}$ cm$^{-2}$.
Eight spectra are generated with the azimuth angle of the corona taken from 0\degree{} to 315\degree{} in increments of 45\degree.
The exposure time for each observation is 40 ks.
The simulated observational data of NuSTAR are analyzed through the lamppost model \texttt{constant\allowbreak \texttimes\allowbreak Tbabs\allowbreak \texttimes\allowbreak relxilllpCp} and the off-axis model \texttt{constant\allowbreak \texttimes\allowbreak Tbabs\allowbreak \texttimes\allowbreak offaxxillCp}.
The energy range 3-79 keV is used in the analysis.

\begin{figure*}[htb!]
    \centering
    \includegraphics[width=0.85\textwidth]{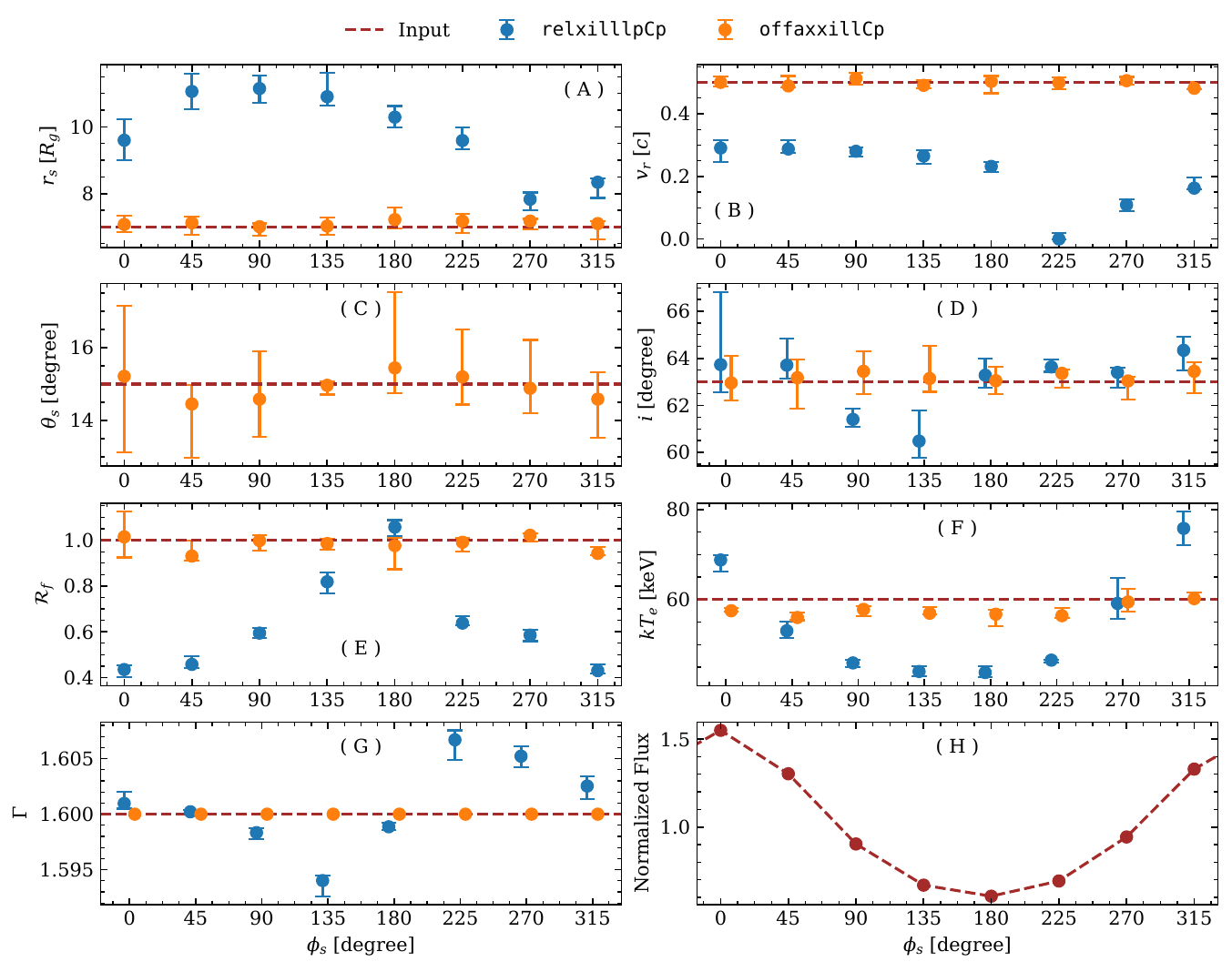}
    \caption{Model parameters measured by \texttt{relxilllpCp} (blue) and \texttt{offaxxillCp} (orange) models from observations of NuSTAR. The input value for each parameter is represented by the brown dashed line. The photon index $\Gamma$ in \texttt{offaxxillCp} model is fixed at 1.6 as in subfigure (G). The subfigure (H) shows the normalized flux of the simulated spectra. The uncertainty of each parameter is given at 90\% confidence.}
    \label{fit}
\end{figure*}

Figure \ref{fit} shows the best-fit values of the variable parameters in two models from observations of NuSTAR. Other parameters are fixed to their input value during fitting.
Both the lamppost model and the off-axis corona model obtained good fitting results for the simulated data, which means the parameters of the model are degenerate.
Parameters measured by the lamppost model are deviated from their input values and exhibit periodic modulations with the azimuth angle of the corona due to the asymmetric illumination profile of the accretion disk from the corona.
On the other, the parameters are restored to their input values using the off-axis corona model, and the off-axis model provides a slightly better fitting than the lamppost model.

The variation of the observed flux with the azimuth angle of the corona is shown in Figure \ref{fit}(H).
The observed flux is maximum where the corona is moving toward the observer ($\phi_s = 0\degree$) and minimum where the corona is moving away from the observer ($\phi_s = 180\degree$).
This is because the source simulated is in the hard state, with the photon index $\Gamma = 1.6$.
The observed flux is dominated by the continuum component of the corona.
When the azimuth angle of the corona changes, the direct radiation from the corona also changes due to the Doppler beaming effect.
The electron temperature of the corona measured by the lamppost model has a similar trend to the observed flux due to the effects of the Doppler shift from the corona to the observer, causing the observed cutoff energy of the continuum component changing with the movement of the corona.
The reflection fraction measured by the lamppost model shows an opposite trend to the observed flux.
This shows that as the direct radiation from the corona increases, there is no corresponding increase in the reflection radiation from the accretion disk, so the measured reflection fraction decreases.

The photon index measured by the lamppost model varies within 0.01 around the input value.
Under the simplifying assumption of the point source, the photon index is hardly affected by the Doppler effects.
That is because the continuum component of the corona can be approximately described as a cutoff power law.
When the corona moves toward or away from the observer, photons reaching the observer are affected by the Doppler effects, causing the observed energy spectrum to be enhanced or weakened and the cutoff energy to be increased or decreased, while the index of the power law is barely affected.
This conclusion is different for an extended corona, of which the hard part close to the black hole has a small photon index, and the soft part far from the black hole has a large photon index.
As the corona moves, the projection area of the different parts of the corona with respect to the observer are modulated with the azimuth angle of the corona, resulting in the modulation of the observed photon index according to the simulations of radiative transfer \citep{2018ApJ...858...82Y, 2020ApJ...897...27Y}.

The observing inclination and the location and motion of the corona describe the profile of the reflection spectrum.
For a corona moving radially outward, the height of the corona is higher than the input value measured with the lamppost model with the change lagging the observed flux by a quarter cycle, which is same to the result given by fitting a single emission line \citep{2023ApJ...955...53F}.
The velocity of the corona measured by the lamppost model is lower than the input value, between 0.1 and 0.3, except when the azimuth angle of the corona is 225\degree, in which the measurement of the corona is almost stationary.
The observing inclination measured by the lamppost model is close to the input value in most cases except for the azimuth angles of the corona of 90\degree{} and 135\degree.
The observing inclination is strongly dependent on the cutoff energy of the blue peak in the emission line profile.
In most cases, the region of the accretion disk moving toward the observer receives enough illumination from the corona, and the observing inclination is measured accurately.
When the corona is moving at the region where the accretion disk is moving away from the observer, the region where the accretion disk is moving toward the observer receives weak illumination from the corona, so the blue peak at the emission line is weak, and the measured observing inclination is less.

\begin{figure*}[htb!]
    \centering
    \includegraphics[width=0.85\textwidth]{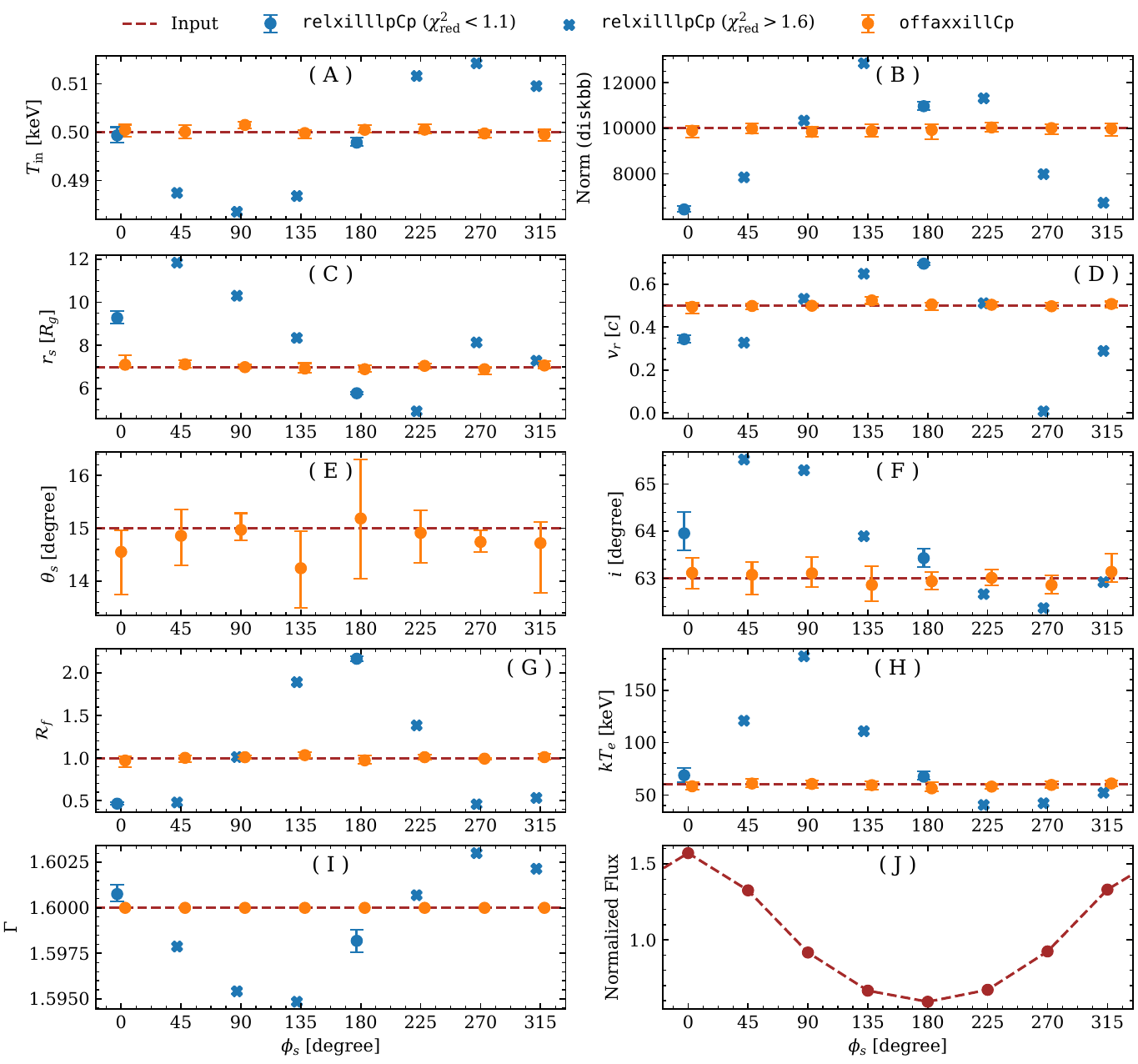}
    \caption{The parameters measured by \texttt{relxilllpCp} (blue) and \texttt{offaxxillCp} (orange) models from observations of eXTP. The legend is the same as in Figure \ref{fit}. For \texttt{relxilllpCp} models with bad fit (reduced chi-square $\chi_\mathrm{red}^2 > 1.6$), the best-fit value of model parameters are represented by blue crosses.}
    \label{fit_eXTP}
\end{figure*}

\begin{figure}[htb!]
    \centering
    \includegraphics[width=0.425\textwidth]{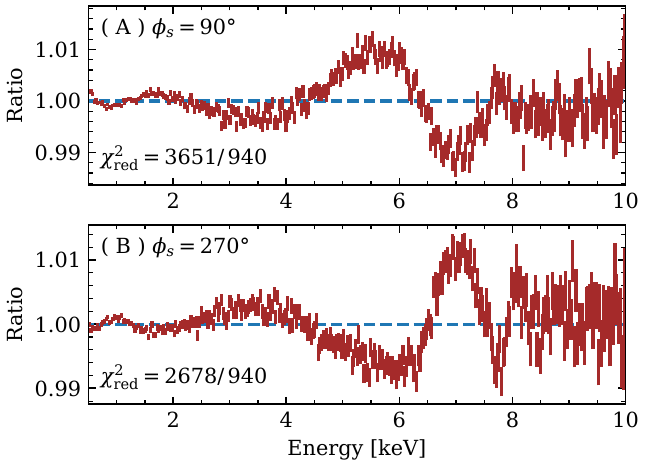}
    \caption{The ratios between simulated data with best-fit \texttt{relxilllpCp} models at the coronal azimuths (A) $\phi_s = 90\degree$ and (B) $\phi_s = 270\degree$. The reduced chi-square statistics are shown at the bottom left of each subfigure.}
    \label{ratio}
\end{figure}

The planned eXTP mission combining high-throughput, good spectral and timing resolution, polarimetric capability, and wide sky coverage will fully unlock the potential of X-ray observations of black holes to study the behavior of matter in strong field gravity \citep{2016SPIE.9905E..1QZ}.
It provides large-area and fast-timing and spectroscopy capability, which allows us to constrain the inner accretion geometry of the black hole \citep{2019SCPMA..6229504D}.
In order to investigate the availability of future high-resolution telescopes to resolve different geometries of corona, we simulate a set of observations with the Spectroscopic Focusing Array (SFA) of eXTP.
The planned SFA is currently under development.
The response file of SFA used in the simulation is version 20241125.
The properties of simulated source are the same as above, except that the blackbody component of the accretion disk is included in the model during the simulation and fitting.
The temperature at inner disk radius $T_\mathrm{in}$ is set to 0.5 keV. The normalization factor is set to $10^4$.
The energy range 0.5-10 keV is used in this analysis.

Figure \ref{fit_eXTP} shows the best-fit values of the variable parameters in the lamppost model \texttt{constant\allowbreak \texttimes\allowbreak Tbabs\allowbreak \texttimes\allowbreak (diskbb\allowbreak +\allowbreak relxilllpCp)} and the off-axis model \texttt{constant\allowbreak \texttimes\allowbreak Tbabs\allowbreak \texttimes\allowbreak (diskbb\allowbreak +\allowbreak offaxxillCp)} from observations of eXTP.
When the azimuth of the corona is at 0\degree{} and 180\degree, the lamppost model obtained a good fit for the simulated data (reduced chi-square $\chi_\mathrm{red}^2 < 1.1$), with no significant components in the residuals.
For other coronal azimuths, it cannot obtain a good fit ($\chi_\mathrm{red}^2 > 1.6$), which is due to the significant redshift or blueshift in the reflection spectra.
The ratios between simulated data with the best-fit lamppost model at the coronal azimuths $\phi_s = 90\degree$ and 270\degree{} are shown in Figure \ref{ratio}.
There are significant residuals at the energy of Fe K$\upalpha$ emission lines.
For the azimuth $\phi_s = 90\degree$, due to the redshift in the emission lines, the flux at the low energy part of the line exceeds the model.
For the azimuth $\phi_s = 270\degree$, the result is opposite due to the blueshift.
When more model parameters are relaxed to free parameters, including the spin of the black hole, the inner disk radius, the ionization, iron abundance, and density of the accretion disk, it still cannot obtain a good fit using the \texttt{relxilllpCp} model.

\section{Discussion}
\label{Discussion}
By fitting the off-axis reflection model to phase-resolved spectra of the QPOs, more investigation of the physical characteristics of the X-ray binaries is possible.
Comparing the simulated observations with the observations of the LFQPOs in MAXI J1820+070 \citep[Figure 6]{2023ApJ...957...84S},
it can be found that our results are consistent in that the dominant contribution to the QPOs comes from the Comptonized component, and the spectral parameters exhibit significant phase modulation.
The electron temperature of the corona in simulated observations shows a similar trend to the observed flux due to the effects of the Doppler shift, which is consistent with the observation in Epochs 3 and 4.
The reflection fraction in simulated observations shows an opposite trend to the observed flux, which is consistent with the observation in Epoch 1 but inconsistent in Epoch 4.
The evolution of the reflection fraction modulation reflects that the accretion geometry of the black hole keeps changing during the outburst.
The observation shows that the photon index varies more than that in the simulation and exhibits the same modulation phase as the QPO flux.
The large variation of the photon index is inconsistent with the assumption of the point source and suggests that the spatial distribution of the corona is complex, such as that assumed in the Lense-Thirring precession model \citep{2009MNRAS.397L.101I} or the jet precession model \citep{2018MNRAS.474L..81L, 2021NatAs...5...94M, 2023ApJ...948..116M}.

In this work, the corona is considered as a point-like source, as in the lamppost scenario, except that it is located off-axis and moves at an arbitrary velocity.
It is necessary to take the spatial dimension of the corona into account before any reliable fitting of the model parameters is performed.
There are many models available for the extended coronal geometry, such as an elongated source moving along the axis of the black hole spin \citep{2013MNRAS.430.1694D}, an extended corona corotating with the accretion disk \citep{2020A&A...641A..89S}, or a disk-like corona above the accretion disk \citep{2022ApJ...925...51R}.
The emissivity produced by the extended corona is similar to a twice-broken power law, with the inner emissivity is determined by the source geometry \citep{2012MNRAS.424.1284W, 2020A&A...641A..89S}.
These emissivity profiles seem to explain some of the observations \citep{2003MNRAS.344L..22M, 2006A&A...453..773S, 2012MNRAS.419..116F}.
However, the simulated observations of black hole binaries show that the properties of the accretion disk and spacetime are generally well recovered even with the lamppost geometry.
It is difficult to constrain the geometry of the corona by spectral fitting method because emissivity profiles generated by quite different coronal geometries can be very similar \citep{2017MNRAS.472.1932G}.

Additional important information on the coronal geometry can be provided by polarimetry, which has been recently greatly advanced with the Imaging X-ray Polarimetry Explorer \citep[IXPE;][]{2016SPIE.9905E..17W}.
The measurements of the X-ray polarization of active galactic nuclei by IXPE tend to favor a slab or a conical shape geometry for the corona \citep{2023MNRAS.523.4468G, 2023JApA...44...87P, 2023MNRAS.525.4735T}.
This indicates that the corona has to be extended and suggests that the motion of the corona matter is important.
The simulations for several corona geometries show that the X-ray polarization signal from the corona is significantly different depending on the coronal geometry \citep{2022MNRAS.510.3674U, 2022RAA....22h5011Y}.
\cite{2022MNRAS.515.2882Z} study the polarization properties of X-ray radiation emerging from off-axis corona using general relativistic Monte Carlo radiative transfer simulations.
The results indicate that for emission averaged over all orbital phases of the coronal rotation, the polarization properties do not change much between the on-axis and off-axis corona.
In view of the short timescale properties of polarization at different azimuths of the corona, it is necessary to make further extensions to our model.

In order to formulate reliable constraints on the system parameters, it will be necessary to include the effects of the self-illumination (or so-called returning radiation) of the accretion disk.
Detailed studies of the effect of self-illumination on the thermal equilibrium of the blackbody radiation emitted from the disk show that the thermal spectrum in which the self-illumination is taken into account appears like a thermal spectrum without it, although having a slightly higher mass-accretion rate \citep{1976ApJ...208..534C, 2005ApJS..157..335L}.
The effects of self-illumination on the full relativistic reflection spectrum as well as the broadened emission lines are complex.
This is because self-illumination changes the shape of the irradiating spectrum, which is no longer a cutoff power law from the corona.
Several studies using approximate methods on the effect of the self-illumination on the reflection spectrum suggest that it is important for fast rotating black holes and when the corona is compact and close to the black hole \citep{2020MNRAS.498.3302W, 2021ApJ...910...49R, 2023EPJC...83..838R, 2022MNRAS.514.3965D}.
\cite{2024ApJ...976..229M, 2024ApJ...965...66M} calculate the reflection spectrum produced by the returning radiation of the thermal and reflection components as well as by the direct radiation from the corona considering the actual spectrum illuminating the disk.
They showed that the returning radiation can have a strong impact on the reflection spectrum when the inner edge of the disk is very close to the black hole, and the corona illuminates mainly the inner part of the accretion disk and in the case of a low or moderate value of the ionization parameter.
In this case, the current reflection model does not fit the data well, and some parameters are overestimated or underestimated significantly.

The emission line energy modulation in the QPO signal could also be caused by the nonaxisymmetric patterns in the accretion disk, such as the emission from accretion flows with spiral wave structures \citep{2001PASJ...53..189K, 2002MNRAS.332L...1H, 2004ApJ...613..700F}.
The spiral wave structure in the accretion disk produces a rich variety of spectra and light curves and enables the source geometry to be constrained through reverberation techniques.
If only the reflection spectrum is considered, there may be some degeneracy in the model fitting when comparing the spiral pattern in the disk with the off-axis corona.
However, the spiral wave model predicts a strong QPO in the disk blackbody components \citep{2013MNRAS.435..749T}.
Although the disk blackbody component is observed to show variations on the QPO timescale, the Comptonized component generally dominates the QPO variability \citep{2016MNRAS.460.2796S, 2024ApJ...973...59S}.
The observation of the reflection fraction makes it possible to distinguish between the spiral pattern and the off-axis corona.
The modulation of the reflection fraction can be naturally explained by precession of the corona or jet as long as their rotation axis is misaligned with the axis of the black hole spin \citep{2019NewAR..8501524I}.
For the spiral wave in the accretion disk, it is unclear how the reflection fraction would change with QPO phase.

\section{Conclusion}
\label{Conclusion}
We present a numerical method for calculating the emission line profile produced by the illumination of an off-axis point source.
By introducing the lens factor, we calculate the observed flux for the distant observer and provide the estimate of the model error.
We obtain new relativistic reflection models by convolving the line model with the accretion disk reflection model.
The new reflection model is consistent with the existing model in the case of lamppost corona geometry.
Off-axis corona has a obvious impact on observations.
Parameters in the model, including corona height, radial velocity, and reflection fraction, cannot be properly measured through the lamppost model.
On the other hand, each model parameter is well obtained by the off-axis corona model.
This shows that the off-axis corona model is more suitable than the lamppost model to study the space distribution and motion of the corona near the black hole through continuous observation of the X-ray energy spectrum, such as the observation of QPOs.

In the case of asymmetric illumination of the accretion disk by the corona, the distribution of ionization parameters of the accretion disk are expected to become important.
The \texttt{relxill} model allows for gradient distributions of ionization parameters and the material density of the accretion disk.
In our model, these two parameters are assumed to be constant.
The difficulty in introducing the gradient distribution into our model is that our model does not make statistics on the illumination profile of the disk in the process of the calculation of the line profile, and it is difficult to obtain the incident flux on the disk simply.
Through improvements in integration methods, including the application of adaptive integration methods, it is possible to treat the distribution of ionization parameters of the accretion disk consistently.
Furthermore, our model ignores the effect of returning radiation from the accretion disk, which requires the calculation of the emission profile received by each point on the disk surface from other points.
Without the assumption of axial symmetry of the system, this is not easy and requires a computation of a huge convolution table.
However, with some appropriate simplifications, the effect of return radiation is possible to add to our model.

\section*{Acknowledgments}
We would like to thank the anonymous referee for their critical and constructive suggestions, and Y.F.Y. is supported by National Natural Science Foundation of China (grant Nos. 12433008, 12393812), National SKA Program of China (grant No. 2020SKA0120300), and the Strategic Priority Research Program of the Chinese Academy of Sciences (grant No. XDB0550200).
S.N.Z. is supported by the National Key R\&D Program of China (2021YFA0718500).

\section*{ORCID iDs}
\noindent Yuan Feng \orcidlink{0000-0003-0554-9425}\url{https://orcid.org/0000-0003-0554-9425} \\
Ye-Fei Yuan \orcidlink{0000-0002-7330-4756}\url{https://orcid.org/0000-0002-7330-4756} \\
Shuang-Nan Zhang \orcidlink{0000-0001-5586-1017}\url{https://orcid.org/0000-0001-5586-1017}

\bibliography{References}

\begin{thebibliography}{}
\expandafter\ifx\csname natexlab\endcsname\relax\def\natexlab#1{#1}\fi
\providecommand{\url}[1]{\href{#1}{#1}}
\providecommand{\dodoi}[1]{doi:~\href{http://doi.org/#1}{\nolinkurl{#1}}}
\providecommand{\doeprint}[1]{\href{http://ascl.net/#1}{\nolinkurl{http://ascl.net/#1}}}
\providecommand{\doarXiv}[1]{\href{https://arxiv.org/abs/#1}{\nolinkurl{https://arxiv.org/abs/#1}}}

\bibitem[{{Arnaud}(1996)}]{1996ASPC..101...17A}
{Arnaud}, K.~A. 1996, in Astronomical Society of the Pacific Conference Series,
  Vol. 101, Astronomical Data Analysis Software and Systems V, ed. G.~H.
  {Jacoby} \& J.~{Barnes}, 17

\bibitem[{{Atri} {et~al.}(2020){Atri}, {Miller-Jones}, {Bahramian}, {Plotkin},
  {Deller}, {Jonker}, {Maccarone}, {Sivakoff}, {Soria}, {Altamirano},
  {Belloni}, {Fender}, {Koerding}, {Maitra}, {Markoff}, {Migliari}, {Russell},
  {Russell}, {Sarazin}, {Tetarenko}, \& {Tudose}}]{2020MNRAS.493L..81A}
{Atri}, P., {Miller-Jones}, J.~C.~A., {Bahramian}, A., {et~al.} 2020, \mnras,
  493, L81, \dodoi{10.1093/mnrasl/slaa010}

\bibitem[{{Bardeen} {et~al.}(1972){Bardeen}, {Press}, \&
  {Teukolsky}}]{1972ApJ...178..347B}
{Bardeen}, J.~M., {Press}, W.~H., \& {Teukolsky}, S.~A. 1972, \apj, 178, 347,
  \dodoi{10.1086/151796}

\bibitem[{{Bennett} {et~al.}(2003){Bennett}, {Halpern}, {Hinshaw}, {Jarosik},
  {Kogut}, {Limon}, {Meyer}, {Page}, {Spergel}, {Tucker}, {Wollack}, {Wright},
  {Barnes}, {Greason}, {Hill}, {Komatsu}, {Nolta}, {Odegard}, {Peiris},
  {Verde}, \& {Weiland}}]{2003ApJS..148....1B}
{Bennett}, C.~L., {Halpern}, M., {Hinshaw}, G., {et~al.} 2003, \apjs, 148, 1,
  \dodoi{10.1086/377253}

\bibitem[{{C{\'a}rdenas-Avenda{\~n}o}
  {et~al.}(2020){C{\'a}rdenas-Avenda{\~n}o}, {Zhou}, \&
  {Bambi}}]{2020PhRvD.101l3014C}
{C{\'a}rdenas-Avenda{\~n}o}, A., {Zhou}, M., \& {Bambi}, C. 2020, \prd, 101,
  123014, \dodoi{10.1103/PhysRevD.101.123014}

\bibitem[{{Cunningham}(1976)}]{1976ApJ...208..534C}
{Cunningham}, C. 1976, \apj, 208, 534, \dodoi{10.1086/154636}

\bibitem[{{Cunningham}(1975)}]{1975ApJ...202..788C}
{Cunningham}, C.~T. 1975, \apj, 202, 788, \dodoi{10.1086/154033}

\bibitem[{{Cunningham} \& {Bardeen}(1973)}]{1973ApJ...183..237C}
{Cunningham}, C.~T., \& {Bardeen}, J.~M. 1973, \apj, 183, 237,
  \dodoi{10.1086/152223}

\bibitem[{{Dauser} {et~al.}(2013){Dauser}, {Garcia}, {Wilms}, {B{\"o}ck},
  {Brenneman}, {Falanga}, {Fukumura}, \& {Reynolds}}]{2013MNRAS.430.1694D}
{Dauser}, T., {Garcia}, J., {Wilms}, J., {et~al.} 2013, \mnras, 430, 1694,
  \dodoi{10.1093/mnras/sts710}

\bibitem[{{Dauser} {et~al.}(2022){Dauser}, {Garc{\'\i}a}, {Joyce},
  {Licklederer}, {Connors}, {Ingram}, {Reynolds}, \&
  {Wilms}}]{2022MNRAS.514.3965D}
{Dauser}, T., {Garc{\'\i}a}, J.~A., {Joyce}, A., {et~al.} 2022, \mnras, 514,
  3965, \dodoi{10.1093/mnras/stac1593}

\bibitem[{{Dauser} {et~al.}(2010){Dauser}, {Wilms}, {Reynolds}, \&
  {Brenneman}}]{2010MNRAS.409.1534D}
{Dauser}, T., {Wilms}, J., {Reynolds}, C.~S., \& {Brenneman}, L.~W. 2010,
  \mnras, 409, 1534, \dodoi{10.1111/j.1365-2966.2010.17393.x}

\bibitem[{{Dauser} {et~al.}(2012){Dauser}, {Svoboda}, {Schartel}, {Wilms},
  {Dov{\v{c}}iak}, {Ehle}, {Karas}, {Santos-Lle{\'o}}, \&
  {Marshall}}]{2012MNRAS.422.1914D}
{Dauser}, T., {Svoboda}, J., {Schartel}, N., {et~al.} 2012, \mnras, 422, 1914,
  \dodoi{10.1111/j.1365-2966.2011.20356.x}

\bibitem[{{De Rosa} {et~al.}(2019){De Rosa}, {Uttley}, {Gou}, {Liu}, {Bambi},
  {Barret}, {Belloni}, {Berti}, {Bianchi}, {Caiazzo}, {Casella}, {Feroci},
  {Ferrari}, {Gualtieri}, {Heyl}, {Ingram}, {Karas}, {Lu}, {Luo}, {Matt},
  {Motta}, {Neilsen}, {Pani}, {Santangelo}, {Shu}, {Wang}, {Wang}, {Xue}, {Xu},
  {Yuan}, {Yuan}, {Zhang}, {Zhang}, {Agudo}, {Amati}, {Andersson}, {Baglio},
  {Bakala}, {Baykal}, {Bhattacharyya}, {Bombaci}, {Bucciantini}, {Capitanio},
  {Ciolfi}, {Cui}, {D'Ammando}, {Dauser}, {Del Santo}, {De Marco}, {Di Salvo},
  {Done}, {Dov{\v{c}}iak}, {Fabian}, {Falanga}, {Gambino}, {Gendre},
  {Grinberg}, {Heger}, {Homan}, {Iaria}, {Jiang}, {Jin}, {Koerding}, {Linares},
  {Liu}, {Maccarone}, {Malzac}, {Manousakis}, {Marin}, {Marinucci},
  {Mehdipour}, {M{\'e}ndez}, {Migliari}, {Miller}, {Miniutti}, {Nardini},
  {O'Brien}, {Osborne}, {Petrucci}, {Possenti}, {Riggio}, {Rodriguez}, {Sanna},
  {Shao}, {Sobolewska}, {Sramkova}, {Stevens}, {Stiele}, {Stratta}, {Stuchlik},
  {Svoboda}, {Tamburini}, {Tauris}, {Tombesi}, {Torok}, {Urbanec}, {Vincent},
  {Wu}, {Yuan}, {in't Zand}, {Zdziarski}, \& {Zhou}}]{2019SCPMA..6229504D}
{De Rosa}, A., {Uttley}, P., {Gou}, L., {et~al.} 2019, Science China Physics,
  Mechanics, and Astronomy, 62, 29504, \dodoi{10.1007/s11433-018-9297-0}

\bibitem[{{Dexter} \& {Agol}(2009)}]{2009ApJ...696.1616D}
{Dexter}, J., \& {Agol}, E. 2009, \apj, 696, 1616,
  \dodoi{10.1088/0004-637X/696/2/1616}

\bibitem[{{Dov{\v{c}}iak} {et~al.}(2004){Dov{\v{c}}iak}, {Karas}, \&
  {Yaqoob}}]{2004ApJS..153..205D}
{Dov{\v{c}}iak}, M., {Karas}, V., \& {Yaqoob}, T. 2004, \apjs, 153, 205,
  \dodoi{10.1086/421115}

\bibitem[{{Draghis} {et~al.}(2023){Draghis}, {Miller}, {Zoghbi}, {Reynolds},
  {Costantini}, {Gallo}, \& {Tomsick}}]{2023ApJ...946...19D}
{Draghis}, P.~A., {Miller}, J.~M., {Zoghbi}, A., {et~al.} 2023, \apj, 946, 19,
  \dodoi{10.3847/1538-4357/acafe7}

\bibitem[{{Fabian} {et~al.}(2000){Fabian}, {Iwasawa}, {Reynolds}, \&
  {Young}}]{2000PASP..112.1145F}
{Fabian}, A.~C., {Iwasawa}, K., {Reynolds}, C.~S., \& {Young}, A.~J. 2000,
  \pasp, 112, 1145, \dodoi{10.1086/316610}

\bibitem[{{Fabian} {et~al.}(1995){Fabian}, {Nandra}, {Reynolds}, {Brandt},
  {Otani}, {Tanaka}, {Inoue}, \& {Iwasawa}}]{1995MNRAS.277L..11F}
{Fabian}, A.~C., {Nandra}, K., {Reynolds}, C.~S., {et~al.} 1995, \mnras, 277,
  L11, \dodoi{10.1093/mnras/277.1.L11}

\bibitem[{{Fabian} {et~al.}(1989){Fabian}, {Rees}, {Stella}, \&
  {White}}]{1989MNRAS.238..729F}
{Fabian}, A.~C., {Rees}, M.~J., {Stella}, L., \& {White}, N.~E. 1989, \mnras,
  238, 729, \dodoi{10.1093/mnras/238.3.729}

\bibitem[{{Fabian} {et~al.}(2009){Fabian}, {Zoghbi}, {Ross}, {Uttley}, {Gallo},
  {Brandt}, {Blustin}, {Boller}, {Caballero-Garcia}, {Larsson}, {Miller},
  {Miniutti}, {Ponti}, {Reis}, {Reynolds}, {Tanaka}, \&
  {Young}}]{2009Natur.459..540F}
{Fabian}, A.~C., {Zoghbi}, A., {Ross}, R.~R., {et~al.} 2009, \nat, 459, 540,
  \dodoi{10.1038/nature08007}

\bibitem[{{Fabian} {et~al.}(2012){Fabian}, {Zoghbi}, {Wilkins}, {Dwelly},
  {Uttley}, {Schartel}, {Miniutti}, {Gallo}, {Grupe}, {Komossa}, \&
  {Santos-Lle{\'o}}}]{2012MNRAS.419..116F}
{Fabian}, A.~C., {Zoghbi}, A., {Wilkins}, D., {et~al.} 2012, \mnras, 419, 116,
  \dodoi{10.1111/j.1365-2966.2011.19676.x}

\bibitem[{{Feng} {et~al.}(2023){Feng}, {Yuan}, \&
  {Zhang}}]{2023ApJ...955...53F}
{Feng}, Y., {Yuan}, Y.-F., \& {Zhang}, S.-N. 2023, \apj, 955, 53,
  \dodoi{10.3847/1538-4357/acedff}

\bibitem[{{Fukumura} \& {Tsuruta}(2004)}]{2004ApJ...613..700F}
{Fukumura}, K., \& {Tsuruta}, S. 2004, \apj, 613, 700, \dodoi{10.1086/423312}

\bibitem[{{Garc{\'\i}a} {et~al.}(2013){Garc{\'\i}a}, {Dauser}, {Reynolds},
  {Kallman}, {McClintock}, {Wilms}, \& {Eikmann}}]{2013ApJ...768..146G}
{Garc{\'\i}a}, J., {Dauser}, T., {Reynolds}, C.~S., {et~al.} 2013, \apj, 768,
  146, \dodoi{10.1088/0004-637X/768/2/146}

\bibitem[{{Garc{\'\i}a} \& {Kallman}(2010)}]{2010ApJ...718..695G}
{Garc{\'\i}a}, J., \& {Kallman}, T.~R. 2010, \apj, 718, 695,
  \dodoi{10.1088/0004-637X/718/2/695}

\bibitem[{{Garc{\'\i}a} {et~al.}(2011){Garc{\'\i}a}, {Kallman}, \&
  {Mushotzky}}]{2011ApJ...731..131G}
{Garc{\'\i}a}, J., {Kallman}, T.~R., \& {Mushotzky}, R.~F. 2011, \apj, 731,
  131, \dodoi{10.1088/0004-637X/731/2/131}

\bibitem[{{Garc{\'\i}a} {et~al.}(2014){Garc{\'\i}a}, {Dauser}, {Lohfink},
  {Kallman}, {Steiner}, {McClintock}, {Brenneman}, {Wilms}, {Eikmann},
  {Reynolds}, \& {Tombesi}}]{2014ApJ...782...76G}
{Garc{\'\i}a}, J., {Dauser}, T., {Lohfink}, A., {et~al.} 2014, \apj, 782, 76,
  \dodoi{10.1088/0004-637X/782/2/76}

\bibitem[{{Gianolli} {et~al.}(2023){Gianolli}, {Kim}, {Bianchi},
  {Ag{\'\i}s-Gonz{\'a}lez}, {Madejski}, {Marin}, {Marinucci}, {Matt}, {Middei},
  {Petrucci}, {Soffitta}, {Tagliacozzo}, {Tombesi}, {Ursini}, {Barnouin}, {De
  Rosa}, {Di Gesu}, {Ingram}, {Loktev}, {Panagiotou}, {Podgorny}, {Poutanen},
  {Puccetti}, {Ratheesh}, {Veledina}, {Zhang}, {Agudo}, {Antonelli},
  {Bachetti}, {Baldini}, {Baumgartner}, {Bellazzini}, {Bongiorno}, {Bonino},
  {Brez}, {Bucciantini}, {Capitanio}, {Castellano}, {Cavazzuti}, {Chen},
  {Ciprini}, {Costa}, {Del Monte}, {Di Lalla}, {Di Marco}, {Donnarumma},
  {Doroshenko}, {Dov{\v{c}}iak}, {Ehlert}, {Enoto}, {Evangelista}, {Fabiani},
  {Ferrazzoli}, {Garc{\'\i}a}, {Gunji}, {Heyl}, {Iwakiri}, {Jorstad}, {Kaaret},
  {Karas}, {Kislat}, {Kitaguchi}, {Kolodziejczak}, {Krawczynski}, {La Monaca},
  {Latronico}, {Liodakis}, {Maldera}, {Manfreda}, {Marscher}, {Marshall},
  {Massaro}, {Mitsuishi}, {Mizuno}, {Muleri}, {Negro}, {Ng}, {O'Dell},
  {Omodei}, {Oppedisano}, {Papitto}, {Pavlov}, {Peirson}, {Perri},
  {Pesce-Rollins}, {Pilia}, {Possenti}, {Ramsey}, {Rankin}, {Roberts},
  {Romani}, {Sgr{\`o}}, {Slane}, {Spandre}, {Swartz}, {Tamagawa}, {Tavecchio},
  {Taverna}, {Tawara}, {Tennant}, {Thomas}, {Trois}, {Tsygankov}, {Turolla},
  {Vink}, {Weisskopf}, {Wu}, {Xie}, \& {Zane}}]{2023MNRAS.523.4468G}
{Gianolli}, V.~E., {Kim}, D.~E., {Bianchi}, S., {et~al.} 2023, \mnras, 523,
  4468, \dodoi{10.1093/mnras/stad1697}

\bibitem[{{Gonzalez} {et~al.}(2017){Gonzalez}, {Wilkins}, \&
  {Gallo}}]{2017MNRAS.472.1932G}
{Gonzalez}, A.~G., {Wilkins}, D.~R., \& {Gallo}, L.~C. 2017, \mnras, 472, 1932,
  \dodoi{10.1093/mnras/stx2080}

\bibitem[{{G{\'o}rski} {et~al.}(2005){G{\'o}rski}, {Hivon}, {Banday},
  {Wandelt}, {Hansen}, {Reinecke}, \& {Bartelmann}}]{2005ApJ...622..759G}
{G{\'o}rski}, K.~M., {Hivon}, E., {Banday}, A.~J., {et~al.} 2005, \apj, 622,
  759, \dodoi{10.1086/427976}

\bibitem[{{Guan} {et~al.}(2021){Guan}, {Tao}, {Qu}, {Zhang}, {Zhang}, {Zhang},
  {Ma}, {Ge}, {Song}, {Lu}, {Li}, {Xu}, {Chen}, {Cao}, {Liu}, {Zhang}, {Wang},
  {Chen}, {Bu}, {Cai}, {Chang}, {Chen}, {Chen}, {Chen}, {Cui}, {Du}, {Gao},
  {Gao}, {Gu}, {Guo}, {Han}, {Huang}, {Huo}, {Jia}, {Jiang}, {Jin}, {Kong},
  {Li}, {Li}, {Li}, {Li}, {Li}, {Li}, {Li}, {Li}, {Liang}, {Liao}, {Liu},
  {Liu}, {Liu}, {Liu}, {Lu}, {Luo}, {Luo}, {Ma}, {Meng}, {Nang}, {Nie}, {Ou},
  {Ren}, {Sai}, {Song}, {Sun}, {Tan}, {Wang}, {Wang}, {Wang}, {Wang}, {Wang},
  {Wen}, {Wu}, {Wu}, {Wu}, {Xiao}, {Xiao}, {Xiong}, {Yang}, {Yang}, {Yang},
  {Yang}, {Yi}, {Yin}, {You}, {Zhang}, {Zhang}, {Zhang}, {Zhang}, {Zhang},
  {Zhang}, {Zhang}, {Zhao}, {Zhao}, {Zheng}, {Zheng}, \&
  {Zhou}}]{2021MNRAS.504.2168G}
{Guan}, J., {Tao}, L., {Qu}, J.~L., {et~al.} 2021, \mnras, 504, 2168,
  \dodoi{10.1093/mnras/stab945}

\bibitem[{{Harrison} {et~al.}(2013){Harrison}, {Craig}, {Christensen},
  {Hailey}, {Zhang}, {Boggs}, {Stern}, {Cook}, {Forster}, {Giommi},
  {Grefenstette}, {Kim}, {Kitaguchi}, {Koglin}, {Madsen}, {Mao}, {Miyasaka},
  {Mori}, {Perri}, {Pivovaroff}, {Puccetti}, {Rana}, {Westergaard}, {Willis},
  {Zoglauer}, {An}, {Bachetti}, {Barri{\`e}re}, {Bellm}, {Bhalerao},
  {Brejnholt}, {Fuerst}, {Liebe}, {Markwardt}, {Nynka}, {Vogel}, {Walton},
  {Wik}, {Alexander}, {Cominsky}, {Hornschemeier}, {Hornstrup}, {Kaspi},
  {Madejski}, {Matt}, {Molendi}, {Smith}, {Tomsick}, {Ajello}, {Ballantyne},
  {Balokovi{\'c}}, {Barret}, {Bauer}, {Blandford}, {Brandt}, {Brenneman},
  {Chiang}, {Chakrabarty}, {Chenevez}, {Comastri}, {Dufour}, {Elvis}, {Fabian},
  {Farrah}, {Fryer}, {Gotthelf}, {Grindlay}, {Helfand}, {Krivonos}, {Meier},
  {Miller}, {Natalucci}, {Ogle}, {Ofek}, {Ptak}, {Reynolds}, {Rigby},
  {Tagliaferri}, {Thorsett}, {Treister}, \& {Urry}}]{2013ApJ...770..103H}
{Harrison}, F.~A., {Craig}, W.~W., {Christensen}, F.~E., {et~al.} 2013, \apj,
  770, 103, \dodoi{10.1088/0004-637X/770/2/103}

\bibitem[{{Hartnoll} \& {Blackman}(2002)}]{2002MNRAS.332L...1H}
{Hartnoll}, S.~A., \& {Blackman}, E.~G. 2002, \mnras, 332, L1,
  \dodoi{10.1046/j.1365-8711.2002.05405.x}

\bibitem[{{Heil} {et~al.}(2015){Heil}, {Uttley}, \&
  {Klein-Wolt}}]{2015MNRAS.448.3348H}
{Heil}, L.~M., {Uttley}, P., \& {Klein-Wolt}, M. 2015, \mnras, 448, 3348,
  \dodoi{10.1093/mnras/stv240}

\bibitem[{{Ingram} \& {Done}(2012)}]{2012MNRAS.427..934I}
{Ingram}, A., \& {Done}, C. 2012, \mnras, 427, 934,
  \dodoi{10.1111/j.1365-2966.2012.21907.x}

\bibitem[{{Ingram} {et~al.}(2009){Ingram}, {Done}, \&
  {Fragile}}]{2009MNRAS.397L.101I}
{Ingram}, A., {Done}, C., \& {Fragile}, P.~C. 2009, \mnras, 397, L101,
  \dodoi{10.1111/j.1745-3933.2009.00693.x}

\bibitem[{{Ingram} {et~al.}(2017){Ingram}, {van der Klis}, {Middleton},
  {Altamirano}, \& {Uttley}}]{2017MNRAS.464.2979I}
{Ingram}, A., {van der Klis}, M., {Middleton}, M., {Altamirano}, D., \&
  {Uttley}, P. 2017, \mnras, 464, 2979, \dodoi{10.1093/mnras/stw2581}

\bibitem[{{Ingram} {et~al.}(2016){Ingram}, {van der Klis}, {Middleton}, {Done},
  {Altamirano}, {Heil}, {Uttley}, \& {Axelsson}}]{2016MNRAS.461.1967I}
{Ingram}, A., {van der Klis}, M., {Middleton}, M., {et~al.} 2016, \mnras, 461,
  1967, \dodoi{10.1093/mnras/stw1245}

\bibitem[{{Ingram} \& {Motta}(2019)}]{2019NewAR..8501524I}
{Ingram}, A.~R., \& {Motta}, S.~E. 2019, \nar, 85, 101524,
  \dodoi{10.1016/j.newar.2020.101524}

\bibitem[{{Kara} {et~al.}(2019){Kara}, {Steiner}, {Fabian}, {Cackett},
  {Uttley}, {Remillard}, {Gendreau}, {Arzoumanian}, {Altamirano}, {Eikenberry},
  {Enoto}, {Homan}, {Neilsen}, \& {Stevens}}]{2019Natur.565..198K}
{Kara}, E., {Steiner}, J.~F., {Fabian}, A.~C., {et~al.} 2019, \nat, 565, 198,
  \dodoi{10.1038/s41586-018-0803-x}

\bibitem[{{Karas} {et~al.}(2001){Karas}, {Martocchia}, \&
  {Subr}}]{2001PASJ...53..189K}
{Karas}, V., {Martocchia}, A., \& {Subr}, L. 2001, \pasj, 53, 189,
  \dodoi{10.1093/pasj/53.2.189}

\bibitem[{{Li} {et~al.}(2005){Li}, {Zimmerman}, {Narayan}, \&
  {McClintock}}]{2005ApJS..157..335L}
{Li}, L.-X., {Zimmerman}, E.~R., {Narayan}, R., \& {McClintock}, J.~E. 2005,
  \apjs, 157, 335, \dodoi{10.1086/428089}

\bibitem[{{Lindquist}(1966)}]{1966AnPhy..37..487L}
{Lindquist}, R.~W. 1966, Annals of Physics, 37, 487,
  \dodoi{10.1016/0003-4916(66)90207-7}

\bibitem[{{Liska} {et~al.}(2018){Liska}, {Hesp}, {Tchekhovskoy}, {Ingram}, {van
  der Klis}, \& {Markoff}}]{2018MNRAS.474L..81L}
{Liska}, M., {Hesp}, C., {Tchekhovskoy}, A., {et~al.} 2018, \mnras, 474, L81,
  \dodoi{10.1093/mnrasl/slx174}

\bibitem[{{Ma} {et~al.}(2021){Ma}, {Tao}, {Zhang}, {Zhang}, {Bu}, {Ge}, {Chen},
  {Qu}, {Zhang}, {Lu}, {Song}, {Yang}, {Yuan}, {Cai}, {Cao}, {Chang}, {Chen},
  {Chen}, {Chen}, {Chen}, {Chen}, {Cui}, {Cui}, {Deng}, {Dong}, {Du}, {Fu},
  {Gao}, {Gao}, {Gao}, {Gu}, {Guan}, {Guo}, {Han}, {Huang}, {Huo}, {Ji}, {Jia},
  {Jiang}, {Jiang}, {Jin}, {Jin}, {Kong}, {Li}, {Li}, {Li}, {Li}, {Li}, {Li},
  {Li}, {Li}, {Li}, {Li}, {Li}, {Liang}, {Liao}, {Liu}, {Liu}, {Liu}, {Liu},
  {Liu}, {Liu}, {Lu}, {Lu}, {Luo}, {Luo}, {Meng}, {Nang}, {Nie}, {Ou}, {Sai},
  {Shang}, {Song}, {Sun}, {Tan}, {Tuo}, {Wang}, {Wang}, {Wang}, {Wang}, {Wang},
  {Wang}, {Wen}, {Wu}, {Wu}, {Wu}, {Xiao}, {Xiao}, {Xie}, {Xiong}, {Xu}, {Xu},
  {Yang}, {Yang}, {Yang}, {Yi}, {Yin}, {You}, {Zhang}, {Zhang}, {Zhang},
  {Zhang}, {Zhang}, {Zhang}, {Zhang}, {Zhang}, {Zhang}, {Zhang}, {Zhang},
  {Zhang}, {Zhang}, {Zhang}, {Zhang}, {Zhang}, {Zhao}, {Zhao}, {Zheng}, {Zhou},
  {Zhou}, {Zhu}, {Zhu}, \& {Zhuang}}]{2021NatAs...5...94M}
{Ma}, X., {Tao}, L., {Zhang}, S.-N., {et~al.} 2021, Nature Astronomy, 5, 94,
  \dodoi{10.1038/s41550-020-1192-2}

\bibitem[{{Ma} {et~al.}(2023){Ma}, {Zhang}, {Tao}, {Bu}, {Qu}, {Zhang}, {Zhou},
  {Huang}, {Jia}, {Song}, {Zhang}, {Ge}, {Liu}, {Yang}, {Yu}, \&
  {Yorgancioglu}}]{2023ApJ...948..116M}
{Ma}, X., {Zhang}, L., {Tao}, L., {et~al.} 2023, \apj, 948, 116,
  \dodoi{10.3847/1538-4357/acc4c3}

\bibitem[{{Martocchia} \& {Matt}(1996)}]{1996MNRAS.282L..53M}
{Martocchia}, A., \& {Matt}, G. 1996, \mnras, 282, L53,
  \dodoi{10.1093/mnras/282.4.L53}

\bibitem[{{Matt} {et~al.}(1991){Matt}, {Perola}, \&
  {Piro}}]{1991A&A...247...25M}
{Matt}, G., {Perola}, G.~C., \& {Piro}, L. 1991, \aap, 247, 25

\bibitem[{{Miller} {et~al.}(2013){Miller}, {Parker}, {Fuerst}, {Bachetti},
  {Harrison}, {Barret}, {Boggs}, {Chakrabarty}, {Christensen}, {Craig},
  {Fabian}, {Grefenstette}, {Hailey}, {King}, {Stern}, {Tomsick}, {Walton}, \&
  {Zhang}}]{2013ApJ...775L..45M}
{Miller}, J.~M., {Parker}, M.~L., {Fuerst}, F., {et~al.} 2013, \apjl, 775, L45,
  \dodoi{10.1088/2041-8205/775/2/L45}

\bibitem[{{Miniutti} {et~al.}(2003){Miniutti}, {Fabian}, {Goyder}, \&
  {Lasenby}}]{2003MNRAS.344L..22M}
{Miniutti}, G., {Fabian}, A.~C., {Goyder}, R., \& {Lasenby}, A.~N. 2003,
  \mnras, 344, L22, \dodoi{10.1046/j.1365-8711.2003.06988.x}

\bibitem[{{Mirzaev} {et~al.}(2024{\natexlab{a}}){Mirzaev}, {Bambi},
  {Abdikamalov}, {Jiang}, {Liu}, {Riaz}, \& {Shashank}}]{2024ApJ...976..229M}
{Mirzaev}, T., {Bambi}, C., {Abdikamalov}, A.~B., {et~al.} 2024{\natexlab{a}},
  \apj, 976, 229, \dodoi{10.3847/1538-4357/ad8a63}

\bibitem[{{Mirzaev} {et~al.}(2024{\natexlab{b}}){Mirzaev}, {Riaz},
  {Abdikamalov}, {Bambi}, {Dauser}, {Garcia}, {Jiang}, {Liu}, \&
  {Shashank}}]{2024ApJ...965...66M}
{Mirzaev}, T., {Riaz}, S., {Abdikamalov}, A.~B., {et~al.} 2024{\natexlab{b}},
  \apj, 965, 66, \dodoi{10.3847/1538-4357/ad303b}

\bibitem[{{Motta} {et~al.}(2015){Motta}, {Casella}, {Henze},
  {Mu{\~n}oz-Darias}, {Sanna}, {Fender}, \& {Belloni}}]{2015MNRAS.447.2059M}
{Motta}, S.~E., {Casella}, P., {Henze}, M., {et~al.} 2015, \mnras, 447, 2059,
  \dodoi{10.1093/mnras/stu2579}

\bibitem[{{Nathan} {et~al.}(2022){Nathan}, {Ingram}, {Homan}, {Huppenkothen},
  {Uttley}, {van der Klis}, {Motta}, {Altamirano}, \&
  {Middleton}}]{2022MNRAS.511..255N}
{Nathan}, E., {Ingram}, A., {Homan}, J., {et~al.} 2022, \mnras, 511, 255,
  \dodoi{10.1093/mnras/stab3803}

\bibitem[{{Pal} {et~al.}(2023){Pal}, {Stalin}, {Chatterjee}, \&
  {Agrawal}}]{2023JApA...44...87P}
{Pal}, I., {Stalin}, C.~S., {Chatterjee}, R., \& {Agrawal}, V.~K. 2023, Journal
  of Astrophysics and Astronomy, 44, 87, \dodoi{10.1007/s12036-023-09981-5}

\bibitem[{{Planck Collaboration} {et~al.}(2011){Planck Collaboration}, {Ade},
  {Aghanim}, {Arnaud}, {Ashdown}, {Aumont}, {Baccigalupi}, {Baker}, {Balbi},
  {Banday}, {Barreiro}, {Bartlett}, {Battaner}, {Benabed}, {Bennett},
  {Beno{\^\i}t}, {Bernard}, {Bersanelli}, {Bhatia}, {Bock}, {Bonaldi}, {Bond},
  {Borrill}, {Bouchet}, {Bradshaw}, {Bremer}, {Bucher}, {Burigana}, {Butler},
  {Cabella}, {Cantalupo}, {Cappellini}, {Cardoso}, {Carr}, {Casale},
  {Catalano}, {Cay{\'o}n}, {Challinor}, {Chamballu}, {Charra}, {Chary},
  {Chiang}, {Chiang}, {Christensen}, {Clements}, {Colombi}, {Couchot},
  {Coulais}, {Crill}, {Crone}, {Crook}, {Cuttaia}, {Danese}, {D'Arcangelo},
  {Davies}, {Davis}, {de Bernardis}, {de Bruin}, {de Gasperis}, {de Rosa}, {de
  Zotti}, {Delabrouille}, {Delouis}, {D{\'e}sert}, {Dick}, {Dickinson},
  {Dolag}, {Dole}, {Donzelli}, {Dor{\'e}}, {D{\"o}rl}, {Douspis}, {Dupac},
  {Efstathiou}, {En{\ss}lin}, {Eriksen}, {Finelli}, {Foley}, {Forni},
  {Fosalba}, {Frailis}, {Franceschi}, {Freschi}, {Gaier}, {Galeotta},
  {Gallegos}, {Gandolfo}, {Ganga}, {Giard}, {Giardino}, {Gienger},
  {Giraud-H{\'e}raud}, {Gonz{\'a}lez}, {Gonz{\'a}lez-Nuevo}, {G{\'o}rski},
  {Gratton}, {Gregorio}, {Gruppuso}, {Guyot}, {Haissinski}, {Hansen},
  {Harrison}, {Helou}, {Henrot-Versill{\'e}}, {Hern{\'a}ndez-Monteagudo},
  {Herranz}, {Hildebrandt}, {Hivon}, {Hobson}, {Holmes}, {Hornstrup}, {Hovest},
  {Hoyland}, {Huffenberger}, {Jaffe}, {Jagemann}, {Jones}, {Juillet}, {Juvela},
  {Kangaslahti}, {Keih{\"a}nen}, {Keskitalo}, {Kisner}, {Kneissl}, {Knox},
  {Krassenburg}, {Kurki-Suonio}, {Lagache}, {L{\"a}hteenm{\"a}ki}, {Lamarre},
  {Lange}, {Lasenby}, {Laureijs}, {Lawrence}, {Leach}, {Leahy}, {Leonardi},
  {Leroy}, {Lilje}, {Linden-V{\o}rnle}, {L{\'o}pez-Caniego}, {Lowe}, {Lubin},
  {Mac{\'\i}as-P{\'e}rez}, {Maciaszek}, {MacTavish}, {Maffei}, {Maino},
  {Mandolesi}, {Mann}, {Maris}, {Mart{\'\i}nez-Gonz{\'a}lez}, {Masi},
  {Massardi}, {Matarrese}, {Matthai}, {Mazzotta}, {McDonald}, {McGehee},
  {Meinhold}, {Melchiorri}, {Melin}, {Mendes}, {Mennella}, {Mevi},
  {Miniscalco}, {Mitra}, {Miville-Desch{\^e}nes}, {Moneti}, {Montier},
  {Morgante}, {Morisset}, {Mortlock}, {Munshi}, {Murphy}, {Naselsky}, {Natoli},
  {Netterfield}, {N{\o}rgaard-Nielsen}, {Noviello}, {Novikov}, {Novikov},
  {O'Dwyer}, {Ortiz}, {Osborne}, {Osuna}, {Oxborrow}, {Pajot}, {Paladini},
  {Partridge}, {Pasian}, {Passvogel}, {Patanchon}, {Pearson}, {Pearson},
  {Perdereau}, {Perotto}, {Perrotta}, {Piacentini}, \&
  {Piat}}]{2011A&A...536A...1P}
{Planck Collaboration}, {Ade}, P.~A.~R., {Aghanim}, N., {et~al.} 2011, \aap,
  536, A1, \dodoi{10.1051/0004-6361/201116464}

\bibitem[{{Ponti} {et~al.}(2010){Ponti}, {Gallo}, {Fabian}, {Miniutti},
  {Zoghbi}, {Uttley}, {Ross}, {Vasudevan}, {Tanaka}, \&
  {Brandt}}]{2010MNRAS.406.2591P}
{Ponti}, G., {Gallo}, L.~C., {Fabian}, A.~C., {et~al.} 2010, \mnras, 406, 2591,
  \dodoi{10.1111/j.1365-2966.2010.16852.x}

\bibitem[{{Reis} {et~al.}(2008){Reis}, {Fabian}, {Ross}, {Miniutti}, {Miller},
  \& {Reynolds}}]{2008MNRAS.387.1489R}
{Reis}, R.~C., {Fabian}, A.~C., {Ross}, R.~R., {et~al.} 2008, \mnras, 387,
  1489, \dodoi{10.1111/j.1365-2966.2008.13358.x}

\bibitem[{{Reynolds}(2013)}]{2013CQGra..30x4004R}
{Reynolds}, C.~S. 2013, Classical and Quantum Gravity, 30, 244004,
  \dodoi{10.1088/0264-9381/30/24/244004}

\bibitem[{{Reynolds}(2021)}]{2021ARA&A..59..117R}
---. 2021, \araa, 59, 117, \dodoi{10.1146/annurev-astro-112420-035022}

\bibitem[{{Riaz} {et~al.}(2022){Riaz}, {Abdikamalov}, {Ayzenberg}, {Bambi},
  {Wang}, \& {Yu}}]{2022ApJ...925...51R}
{Riaz}, S., {Abdikamalov}, A.~B., {Ayzenberg}, D., {et~al.} 2022, \apj, 925,
  51, \dodoi{10.3847/1538-4357/ac3827}

\bibitem[{{Riaz} {et~al.}(2023){Riaz}, {Mirzaev}, {Abdikamalov}, \&
  {Bambi}}]{2023EPJC...83..838R}
{Riaz}, S., {Mirzaev}, T., {Abdikamalov}, A.~B., \& {Bambi}, C. 2023, European
  Physical Journal C, 83, 838, \dodoi{10.1140/epjc/s10052-023-12031-7}

\bibitem[{{Riaz} {et~al.}(2021){Riaz}, {Szanecki}, {Nied{\'z}wiecki},
  {Ayzenberg}, \& {Bambi}}]{2021ApJ...910...49R}
{Riaz}, S., {Szanecki}, M., {Nied{\'z}wiecki}, A., {Ayzenberg}, D., \& {Bambi},
  C. 2021, \apj, 910, 49, \dodoi{10.3847/1538-4357/abe2a3}

\bibitem[{{Risaliti} {et~al.}(2013){Risaliti}, {Harrison}, {Madsen}, {Walton},
  {Boggs}, {Christensen}, {Craig}, {Grefenstette}, {Hailey}, {Nardini},
  {Stern}, \& {Zhang}}]{2013Natur.494..449R}
{Risaliti}, G., {Harrison}, F.~A., {Madsen}, K.~K., {et~al.} 2013, \nat, 494,
  449, \dodoi{10.1038/nature11938}

\bibitem[{{Ross} \& {Fabian}(2005)}]{2005MNRAS.358..211R}
{Ross}, R.~R., \& {Fabian}, A.~C. 2005, \mnras, 358, 211,
  \dodoi{10.1111/j.1365-2966.2005.08797.x}

\bibitem[{{Ruszkowski}(2000)}]{2000MNRAS.315....1R}
{Ruszkowski}, M. 2000, \mnras, 315, 1, \dodoi{10.1046/j.1365-8711.2000.02898.x}

\bibitem[{{Shui} {et~al.}(2023){Shui}, {Zhang}, {Zhang}, {Chen}, {Kong},
  {Wang}, {Peng}, {Ji}, {Santangelo}, {Yin}, {Qu}, {Tao}, {Ge}, {Huang},
  {Zhang}, {Liu}, {Zhang}, {Yu}, {Chang}, {Li}, {Ye}, {Li}, {Yu}, \&
  {Yan}}]{2023ApJ...957...84S}
{Shui}, Q.~C., {Zhang}, S., {Zhang}, S.~N., {et~al.} 2023, \apj, 957, 84,
  \dodoi{10.3847/1538-4357/acfc42}

\bibitem[{{Shui} {et~al.}(2024){Shui}, {Zhang}, {Peng}, {Zhang}, {Chen}, {Ji},
  {Kong}, {Feng}, {Yu}, {Wang}, {Chang}, {Yin}, {Qu}, {Tao}, {Ge}, {Zhang}, \&
  {Li}}]{2024ApJ...973...59S}
{Shui}, Q.-C., {Zhang}, S., {Peng}, J.-Q., {et~al.} 2024, \apj, 973, 59,
  \dodoi{10.3847/1538-4357/ad676a}

\bibitem[{{Stevens} \& {Uttley}(2016)}]{2016MNRAS.460.2796S}
{Stevens}, A.~L., \& {Uttley}, P. 2016, \mnras, 460, 2796,
  \dodoi{10.1093/mnras/stw1093}

\bibitem[{{Suebsuwong} {et~al.}(2006){Suebsuwong}, {Malzac}, {Jourdain}, \&
  {Marcowith}}]{2006A&A...453..773S}
{Suebsuwong}, T., {Malzac}, J., {Jourdain}, E., \& {Marcowith}, A. 2006, \aap,
  453, 773, \dodoi{10.1051/0004-6361:20064973}

\bibitem[{{Szanecki} {et~al.}(2020){Szanecki}, {Nied{\'z}wiecki}, {Done},
  {Klepczarek}, {Lubi{\'n}ski}, \& {Mizumoto}}]{2020A&A...641A..89S}
{Szanecki}, M., {Nied{\'z}wiecki}, A., {Done}, C., {et~al.} 2020, \aap, 641,
  A89, \dodoi{10.1051/0004-6361/202038303}

\bibitem[{{Tagliacozzo} {et~al.}(2023){Tagliacozzo}, {Marinucci}, {Ursini},
  {Matt}, {Bianchi}, {Baldini}, {Barnouin}, {Cavero Rodriguez}, {De Rosa}, {Di
  Gesu}, {Dov{\v{c}}iak}, {Harper}, {Ingram}, {Karas}, {Kim}, {Krawczynski},
  {Madejski}, {Marin}, {Middei}, {Marshall}, {Muleri}, {Panagiotou},
  {Petrucci}, {Podgorny}, {Poutanen}, {Puccetti}, {Soffitta}, {Tombesi},
  {Veledina}, {Zhang}, {Agudo}, {Antonelli}, {Bachetti}, {Baumgartner},
  {Bellazzini}, {Bongiorno}, {Bonino}, {Brez}, {Bucciantini}, {Capitanio},
  {Castellano}, {Cavazzuti}, {Chen}, {Ciprini}, {Costa}, {Del Monte}, {Di
  Lalla}, {Di Marco}, {Donnarumma}, {Doroshenko}, {Ehlert}, {Enoto},
  {Evangelista}, {Fabiani}, {Ferrazzoli}, {Garcia}, {Gunji}, {Heyl}, {Iwakiri},
  {Jorstad}, {Kaaret}, {Kislat}, {Kitaguchi}, {Kolodziejczak}, {La Monaca},
  {Latronico}, {Liodakis}, {Maldera}, {Manfreda}, {Marscher}, {Massaro},
  {Mitsuishi}, {Mizuno}, {Negro}, {Ng}, {O'Dell}, {Omodei}, {Oppedisano},
  {Papitto}, {Pavlov}, {Peirson}, {Perri}, {Pesce-Rollins}, {Pilia},
  {Possenti}, {Ramsey}, {Rankin}, {Ratheesh}, {Roberts}, {Romani}, {Sgr{\`o}},
  {Slane}, {Spandre}, {Swartz}, {Tamagawa}, {Tavecchio}, {Taverna}, {Tawara},
  {Tennant}, {Thomas}, {Trois}, {Tsygankov}, {Turolla}, {Vink}, {Weisskopf},
  {Wu}, {Xie}, \& {Zane}}]{2023MNRAS.525.4735T}
{Tagliacozzo}, D., {Marinucci}, A., {Ursini}, F., {et~al.} 2023, \mnras, 525,
  4735, \dodoi{10.1093/mnras/stad2627}

\bibitem[{{Tanaka} {et~al.}(1995){Tanaka}, {Nandra}, {Fabian}, {Inoue},
  {Otani}, {Dotani}, {Hayashida}, {Iwasawa}, {Kii}, {Kunieda}, {Makino}, \&
  {Matsuoka}}]{1995Natur.375..659T}
{Tanaka}, Y., {Nandra}, K., {Fabian}, A.~C., {et~al.} 1995, \nat, 375, 659,
  \dodoi{10.1038/375659a0}

\bibitem[{{Tripathi} {et~al.}(2021){Tripathi}, {Abdikamalov}, {Ayzenberg},
  {Bambi}, \& {Liu}}]{2021ApJ...913..129T}
{Tripathi}, A., {Abdikamalov}, A.~B., {Ayzenberg}, D., {Bambi}, C., \& {Liu},
  H. 2021, \apj, 913, 129, \dodoi{10.3847/1538-4357/abf6c5}

\bibitem[{{Tsang} \& {Butsky}(2013)}]{2013MNRAS.435..749T}
{Tsang}, D., \& {Butsky}, I. 2013, \mnras, 435, 749,
  \dodoi{10.1093/mnras/stt1334}

\bibitem[{{Ursini} {et~al.}(2022){Ursini}, {Matt}, {Bianchi}, {Marinucci},
  {Dov{\v{c}}iak}, \& {Zhang}}]{2022MNRAS.510.3674U}
{Ursini}, F., {Matt}, G., {Bianchi}, S., {et~al.} 2022, \mnras, 510, 3674,
  \dodoi{10.1093/mnras/stab3745}

\bibitem[{{van den Eijnden} {et~al.}(2017){van den Eijnden}, {Ingram},
  {Uttley}, {Motta}, {Belloni}, \& {Gardenier}}]{2017MNRAS.464.2643V}
{van den Eijnden}, J., {Ingram}, A., {Uttley}, P., {et~al.} 2017, \mnras, 464,
  2643, \dodoi{10.1093/mnras/stw2634}

\bibitem[{{{\v{C}}ade{\v{z}}} {et~al.}(1998){{\v{C}}ade{\v{z}}}, {Fanton}, \&
  {Calvani}}]{1998NewA....3..647C}
{{\v{C}}ade{\v{z}}}, A., {Fanton}, C., \& {Calvani}, M. 1998, \na, 3, 647,
  \dodoi{10.1016/S1384-1076(98)00035-9}

\bibitem[{{Weisskopf} {et~al.}(2016){Weisskopf}, {Ramsey}, {O'Dell}, {Tennant},
  {Elsner}, {Soffitta}, {Bellazzini}, {Costa}, {Kolodziejczak}, {Kaspi},
  {Muleri}, {Marshall}, {Matt}, \& {Romani}}]{2016SPIE.9905E..17W}
{Weisskopf}, M.~C., {Ramsey}, B., {O'Dell}, S., {et~al.} 2016, in Society of
  Photo-Optical Instrumentation Engineers (SPIE) Conference Series, Vol. 9905,
  Space Telescopes and Instrumentation 2016: Ultraviolet to Gamma Ray, ed.
  J.-W.~A. {den Herder}, T.~{Takahashi}, \& M.~{Bautz}, 990517,
  \dodoi{10.1117/12.2235240}

\bibitem[{{Wilkins} \& {Fabian}(2012)}]{2012MNRAS.424.1284W}
{Wilkins}, D.~R., \& {Fabian}, A.~C. 2012, \mnras, 424, 1284,
  \dodoi{10.1111/j.1365-2966.2012.21308.x}

\bibitem[{{Wilkins} {et~al.}(2020){Wilkins}, {Garc{\'\i}a}, {Dauser}, \&
  {Fabian}}]{2020MNRAS.498.3302W}
{Wilkins}, D.~R., {Garc{\'\i}a}, J.~A., {Dauser}, T., \& {Fabian}, A.~C. 2020,
  \mnras, 498, 3302, \dodoi{10.1093/mnras/staa2566}

\bibitem[{{Wilms} {et~al.}(2001){Wilms}, {Reynolds}, {Begelman}, {Reeves},
  {Molendi}, {Staubert}, \& {Kendziorra}}]{2001MNRAS.328L..27W}
{Wilms}, J., {Reynolds}, C.~S., {Begelman}, M.~C., {et~al.} 2001, \mnras, 328,
  L27, \dodoi{10.1046/j.1365-8711.2001.05066.x}

\bibitem[{{Yang} \& {Wang}(2013)}]{2013ApJS..207....6Y}
{Yang}, X., \& {Wang}, J. 2013, \apjs, 207, 6,
  \dodoi{10.1088/0067-0049/207/1/6}

\bibitem[{{Yang} {et~al.}(2022){Yang}, {Wang}, \& {Yang}}]{2022RAA....22h5011Y}
{Yang}, X.-L., {Wang}, J.-C., \& {Yang}, C.-Y. 2022, Research in Astronomy and
  Astrophysics, 22, 085011, \dodoi{10.1088/1674-4527/ac7543}

\bibitem[{{You} {et~al.}(2018){You}, {Bursa}, \&
  {{\.Z}ycki}}]{2018ApJ...858...82Y}
{You}, B., {Bursa}, M., \& {{\.Z}ycki}, P.~T. 2018, \apj, 858, 82,
  \dodoi{10.3847/1538-4357/aabd33}

\bibitem[{{You} {et~al.}(2020){You}, {{\.Z}ycki}, {Ingram}, {Bursa}, \&
  {Wang}}]{2020ApJ...897...27Y}
{You}, B., {{\.Z}ycki}, P.~T., {Ingram}, A., {Bursa}, M., \& {Wang}, W. 2020,
  \apj, 897, 27, \dodoi{10.3847/1538-4357/ab9838}

\bibitem[{{Yu} \& {Lu}(2000)}]{2000MNRAS.311..161Y}
{Yu}, Q., \& {Lu}, Y. 2000, \mnras, 311, 161,
  \dodoi{10.1046/j.1365-8711.2000.03019.x}

\bibitem[{{Yuan} {et~al.}(2009){Yuan}, {Cao}, {Huang}, \&
  {Shen}}]{2009ApJ...699..722Y}
{Yuan}, Y.-F., {Cao}, X., {Huang}, L., \& {Shen}, Z.-Q. 2009, \apj, 699, 722,
  \dodoi{10.1088/0004-637X/699/1/722}

\bibitem[{{Zhang} {et~al.}(2014){Zhang}, {Lu}, {Zhang}, \&
  {Li}}]{2014SPIE.9144E..21Z}
{Zhang}, S., {Lu}, F.~J., {Zhang}, S.~N., \& {Li}, T.~P. 2014, in Society of
  Photo-Optical Instrumentation Engineers (SPIE) Conference Series, Vol. 9144,
  Space Telescopes and Instrumentation 2014: Ultraviolet to Gamma Ray, ed.
  T.~{Takahashi}, J.-W.~A. {den Herder}, \& M.~{Bautz}, 914421,
  \dodoi{10.1117/12.2054144}

\bibitem[{{Zhang} {et~al.}(2016){Zhang}, {Feroci}, {Santangelo}, {Dong},
  {Feng}, {Lu}, {Nandra}, {Wang}, {Zhang}, {Bozzo}, {Brandt}, {De Rosa}, {Gou},
  {Hernanz}, {van der Klis}, {Li}, {Liu}, {Orleanski}, {Pareschi}, {Pohl},
  {Poutanen}, {Qu}, {Schanne}, {Stella}, {Uttley}, {Watts}, {Xu}, {Yu}, {in 't
  Zand}, {Zane}, {Alvarez}, {Amati}, {Baldini}, {Bambi}, {Basso},
  {Bhattacharyya S.}, {}, {Belloni}, {Bellutti}, {Bianchi}, {Brez}, {Bursa},
  {Burwitz}, {Budtz-J{\o}rgensen}, {Caiazzo}, {Campana}, {Cao}, {Casella},
  {Chen}, {Chen}, {Chen}, {Chen}, {Chen}, {Chen}, {Civitani}, {Coti Zelati},
  {Cui}, {Cui}, {Dai}, {Del Monte}, {de Martino}, {Di Cosimo}, {Diebold},
  {Dovciak}, {Donnarumma}, {Doroshenko}, {Esposito}, {Evangelista}, {Favre},
  {Friedrich}, {Fuschino}, {Galvez}, {Gao}, {Ge}, {Gevin}, {Goetz}, {Han},
  {Heyl}, {Horak}, {Hu}, {Huang}, {Huang}, {Hudec}, {Huppenkothen}, {Israel},
  {Ingram}, {Karas}, {Karelin}, {Jenke}, {Ji}, {Korpela}, {Kunneriath},
  {Labanti}, {Li}, {Li}, {Li}, {Liang}, {Limousin}, {Lin}, {Ling}, {Liu},
  {Liu}, {Liu}, {Lu}, {Lund}, {Lai}, {Luo}, {Luo}, {Ma}, {Mahmoodifar},
  {Marisaldi}, {Martindale}, {Meidinger}, {Men}, {Michalska}, {Mignani},
  {Minuti}, {Motta}, {Muleri}, {Neilsen}, {Orlandini}, {Pan}, {Patruno},
  {Perinati}, {Picciotto}, {Piemonte}, {Pinchera}, {Rachevski A.}, {Rapisarda},
  {Rea}, {Rossi}, {Rubini}, {Sala}, {Shu}, {Sgro}, {Shen}, {Soffitta}, {Song},
  {Spandre}, {Stratta}, {Strohmayer}, {Sun}, {Svoboda}, {Tagliaferri},
  {Tenzer}, {Hong}, {Taverna}, {Torok}, {Turolla}, {Vacchi}, {Wang}, {Walton},
  {Wang}, {Wang}, {Wang}, {Wang}, {Weng}, {Wilms}, {Winter}, {Wu}, {Wu},
  {Xiong}, {Xu}, {Xue}, {Yan}, {Yang}, {Yang}, {Yang}, {Yuan}, {Yuan}, {Yuan},
  {Zampa}, {Zampa}, {Zdziarski}, {Zhang}, {Zhang}, {Zhang}, {Zhang}, {Zhang},
  {Zhang}, {Zheng}, {Zhou}, \& {Zhou X.~L.}}]{2016SPIE.9905E..1QZ}
{Zhang}, S.~N., {Feroci}, M., {Santangelo}, A., {et~al.} 2016, in Society of
  Photo-Optical Instrumentation Engineers (SPIE) Conference Series, Vol. 9905,
  Space Telescopes and Instrumentation 2016: Ultraviolet to Gamma Ray, ed.
  J.-W.~A. {den Herder}, T.~{Takahashi}, \& M.~{Bautz}, 99051Q,
  \dodoi{10.1117/12.2232034}

\bibitem[{{Zhang} {et~al.}(2020){Zhang}, {Li}, {Lu}, {Song}, {Xu}, {Liu},
  {Chen}, {Cao}, {Bu}, {Chang}, {Chen}, {Chen}, {Chen}, {Chen}, {Chen}, {Cui},
  {Cui}, {Deng}, {Dong}, {Du}, {Fu}, {Gao}, {Gao}, {Gao}, {Ge}, {Gu}, {Guan},
  {Gungor}, {Guo}, {Han}, {Hu}, {Huang}, {Huo}, {Jia}, {Jiang}, {Jiang}, {Jin},
  {Jin}, {Li}, {Li}, {Li}, {Li}, {Li}, {Li}, {Li}, {Li}, {Li}, {Li}, {Li},
  {Liang}, {Liao}, {Liu}, {Liu}, {Liu}, {Liu}, {Liu}, {Liu}, {Lu}, {Lu}, {Luo},
  {Ma}, {Meng}, {Nang}, {Nie}, {Ou}, {Qu}, {Sai}, {Shang}, {Shen}, {Sun},
  {Tan}, {Tao}, {Tuo}, {Wang}, {Wang}, {Wang}, {Wang}, {Wang}, {Wang}, {Wang},
  {Wen}, {Wu}, {Wu}, {Wu}, {Xiao}, {Xiong}, {Yan}, {Yang}, {Yang}, {Yang},
  {Yi}, {Yuan}, {Zhang}, {Zhang}, {Zhang}, {Zhang}, {Zhang}, {Zhang}, {Zhang},
  {Zhang}, {Zhang}, {Zhang}, {Zhang}, {Zhang}, {Zhang}, {Zhang}, {Zhang},
  {Zhang}, {Zhang}, {Zhang}, {Zhang}, {Zhang}, {Zhao}, {Zhao}, {Zheng}, {Zhou},
  {Zhu}, {Zhu}, {Zhuang}, \& {Insight-HXMT Team}}]{2020SCPMA..6349502Z}
{Zhang}, S.-N., {Li}, T., {Lu}, F., {et~al.} 2020, Science China Physics,
  Mechanics, and Astronomy, 63, 249502, \dodoi{10.1007/s11433-019-1432-6}

\bibitem[{{Zhang} {et~al.}(2022){Zhang}, {Dov{\v{c}}iak}, {Bursa}, {Karas},
  {Matt}, \& {Ursini}}]{2022MNRAS.515.2882Z}
{Zhang}, W., {Dov{\v{c}}iak}, M., {Bursa}, M., {et~al.} 2022, \mnras, 515,
  2882, \dodoi{10.1093/mnras/stac1937}

\bibitem[{{Zhao} {et~al.}(2021){Zhao}, {Gou}, {Dong}, {Tuo}, {Liao}, {Li},
  {Jia}, {Feng}, \& {Steiner}}]{2021ApJ...916..108Z}
{Zhao}, X., {Gou}, L., {Dong}, Y., {et~al.} 2021, \apj, 916, 108,
  \dodoi{10.3847/1538-4357/ac07a9}

\end{thebibliography}
\end{document}